\documentclass[11pt]{iopart}
\usepackage{iopams}
\usepackage{epsf}
\def\r{\mathbb R}                   
\def\NAT{\mathbb N}                   
\def\z{\mathbb Z}                   
\def\L{\mathbb L}
\def\E{\mathbb E}
\def\RW{\mathbb R\mathbb W}
\def\AdS{\mathbb{A}d\mathbb{S}}
\def\dS{d\mathbb{S}}
\def\Sc{\mathbb{S}}
\def\RN{\mathbb R\mathbb N}
\def\ppW{pp\mathbb W}
\def\PW{\mathbb P\mathbb W}
\def\BI{\mathbb{B}_{{\rm I}}}
\def\K{\mathbb K}
\def\d{\partial}
\def\fr{\frac}
\def\be{\begin{equation}}
\def\ee{\end{equation}}
\def\bea{\begin{eqnarray}}
\def\eea{\end{eqnarray}}
\def\e*{\end{eqnarray*}}
\def\N{\hfill \raisebox{1mm}{\framebox{\rule{0mm}{1mm}}}}
\def\T{\tilde{{\bf g}}}

\def\S{\Sigma}

\def\G{{\bf g}}
\def\gg{\mbox{g}}
\def\V{\it V}
\def\n{\it n}
\def\DP{{\cal DP}}

\def\g{\gamma}

\def\Z{\Theta}
\def\z{\theta}
\def\O{\Omega}
\def\p{\Psi}
\def\f{\varphi}
\def\x{\chi}
\def\P{{\it Proof :} \hspace{3mm}}
\def\X{\vec{X}}
\def\Y{\vec{Y}}
\def\k{\vec{k}}

\def\W{{\cal W}}
\def\B{{\cal B}}
\def\U{{\cal U}}
\def\NN{{\cal N}}

\def\l{\lambda}

\def\u{\vec u}

\def\t{\tau}
\newtheorem{defi}{Definition}[section]
\newtheorem{theo}{Theorem}[section]
\newtheorem{coro}{Corollary}[section]
\newtheorem{prop}{Proposition}[section]
\newtheorem{lem}{Lemma}[section]

\newtheorem{cri}{Criterion}
\begin{document}

\title[Causal Relationship: a new tool for the causal
characterization of ...]
{Causal Relationship: a new tool for the causal
characterization of Lorentzian manifolds}

\author{Alfonso Garc\'{\i}a-Parrado and Jos\'{e} M M Senovilla}

\address{Departamento de F\'{\i}sica Te\'{o}rica,
Universidad del Pa\'{\i}s Vasco, Apartado 644, 48080 Bilbao, Spain}

\eads{\mailto{wtbgagoa@lg.ehu.es} and \mailto{wtpmasej@lg.ehu.es}}

\begin{abstract}
We define and study a new kind of relation between two
diffeomorphic Lorentzian manifolds called {\em causal relation}, which
is any diffeomorphism characterized by mapping every causal
vector of the first manifold onto a causal vector of the second.
We perform a thorough study of the
mathematical properties of causal relations and prove in particular
that two given Lorentzian manifolds (say $V$ and $W$)
may be causally related only in one direction (say from $V$ to $W$,
but not from $W$ to $V$).  This leads us to the concept of
causally equivalent (or {\em isocausal} in short) Lorentzian manifolds
as those mutually causally related and to a definition of {\em causal structure}
over a differentiable manifold as the equivalence class formed by
isocausal Lorentzian metrics upon it. 

Isocausality is a more general concept than
the conformal relationship, because we prove
the remarkable result that a conformal relation $\f$ is characterized
by the fact of being a causal relation of the {\em particular} kind
in which both $\f$ and $\f^{-1}$ are causal relations. 
Isocausal Lorentzian manifolds are mutually causally compatible, they 
share some important causal properties, and there are
one-to-one correspondences, which sometimes are non-trivial,
between several classes of their respective future (and past) objects.
A more important feature is that they
satisfy the same standard  causality constraints.
We also introduce a partial order for the
equivalence classes of isocausal Lorentzian manifolds providing
a classification of all the
causal structures that a given fixed manifold can have.

By introducing the concept of {\em causal extension} we put forward a new
definition of {\em causal boundary} for Lorentzian manifolds
based on the concept of isocausality, and thereby we generalize the
traditional Penrose's constructions of conformal infinity, diagrams and
embeddings. In particular, the concept of {\em causal diagram} is given.

Many explicit clarifying examples are presented throughout the
paper.
\end{abstract}

\pacs{04.20.Cv, 04.20.Gz, 02.40.-k}

\submitto{\CQG}

\section{Introduction}
Causality is one of the most important and basic concepts in physical
theories,
and in particular in all relativistic theories based on a Lorentzian
manifold,
including the outstanding cases of General Relativity and its
relatives.
 From the mathematical viewpoint, causality is an essential
ingredient,
genuine to the Lorentzian structure of the spacetime, which lies at a
level
above the underlying manifold arena but below the more complete metric
structure.
 From a physical point of view, it embodies the concepts of time
evolution,
finite speed of signal propagation, and accessible communications.
 Furthermore,
causality concepts were crucial in many important achievements
of gravitational theories, such as the singularity theorems (see e.g.\
\cite{FF,PENROSE,BEE,COND}), the study of initial value formulations
(e.g. \cite{ChG,FF}), or the definitions of asymptotic properties of
the spacetime
(e.g. \cite{P1,P2,FF}). Several types of causality conditions
\cite{Carter,FF,COND} are usually required
on spacetimes in order to ensure their physical reasonability.
Our main goal in this paper is to introduce a new tool which may be 
helpful in the characterization of the causal structure of spacetimes 
and is compatible with those causality conditions.

Conventional wisdom states that the causal structure of spacetimes is
given by its conformal structure, that is to say, two spacetimes have 
identical causality properties if they are
related by a conformal diffeomorphism. This is based on the fact 
that, as is well known, conformal relations map causal vectors onto causal
vectors preserving their causal character (i.e.\ null vectors
are mapped to null vectors and timelike vectors to timelike ones).
Another way of putting it is that such a causal structure
determines the metric of the spacetime up to a conformal factor. This view is
acceptable, as conformal mappings keep the causal
properties of spacetimes {\em faithfully}, but it is neither unique
nor fully general.
In particular, we will see that another possibility for the concept of
{\em global causal equivalence} between Lorentzian manifolds,
keeping many important causal properties, can be defined.
This new definition of causal equivalence, its motivations, and the 
concept of causal structure derived from it ---which is more general 
than the conformal one--- will be one of the main subjects of this work.

In order to carry out our program we will remove the
restriction that the causal character of the vectors must be
preserved. Thus, a {\em causal relation}---also called a {\em causal
mapping} --- will be a global diffeomorphism that simply maps
future-directed vectors onto future-directed vectors. This idea has
been considered previously, specially in connection with a possible
theory of gravity on a flat background, see
e.g.\cite{PS1,PS2} and references therein. In
such special relativistic theories of gravity one needs that the
causality defined by the curved metric tensor be compatible with that of the
underlying flat background metric \cite{PS2,P3}, and this
compatibility can be properly defined by the causal mapping.
Nevertheless, our approach will be of complete general nature and not
related to any, flat or not, fixed background. Moreover, we will
go one step further: we will consider the {\it mutual causal
compatibility}, that is, the existence of reciprocal causal relations
between two Lorentzian manifolds. As we will show, there exist situations
in which we
can set up a causal mapping from one spacetime to another, but not
the other way round. The point here is that, as opposed to what happens with
conformal relations, the inverse diffeomorphism is not necessarily a causal
relation.
However, there may be {\em other} diffeomorphisms in the inverse
direction which are certainly causal relations.
Therefore it makes sense to define {\em isocausal Lorentzian
manifolds} as those for which it is possible to establish the causal
relations in both ways. Of course globally conformally related Lorentzian
manifolds are isocausal but the converse does not hold and there are
many isocausal spacetimes which are not globally conformal; we will
exhibit explicit examples. Actually, we will identify the conformal
relations as the unique causal relations whose inverse diffeomorphism
is also a causal relation. An important number of causal properties,
as well as the standard causality constraints, remain the same for
isocausal spacetimes rendering isocausality as a good definition that
generalizes the conformal equivalence.

The equivalence classes formed by isocausal spacetimes---which we call
causal structures--- can be naturally
partially ordered in a way which will be shown to be a
refinement and an improvement of the well-known causality-condition hierarchy
in General Relativity. Therefore, isocausality is a concept stronger 
than the standard causal hierarchy but weaker than the conformal 
equivalence. This intermediate position will allow to do some things 
not permitted by the conformal relationship while keeping an 
important part of the causal properties of spacetimes. The usefulness 
of this intermediate position will be analyzed in some examples along 
this paper. We believe that isocausality may be helpful in 
understanding the causal properties shared by some non-conformally 
related spacetimes. Furthermore, the concept of {\em partly conformal} 
spacetimes, in the sense of having subspaces which are truly conformal,
can be consistently defined by means of isocausality, and 
this may have important applications in cases of physical interest 
such as decomposable metrics, warped products, spherical symmetry, and 
so on, as well as in the generalization of Penrose diagrams, see below.
It is also possible to construct ordered sequences of isocausal spacetimes,
or of causal structures in a manifold, in such a way
that the worst (best) causally behaved spacetimes are the greatest
(lowest)
elements of the sequence. This raises the interesting question as to
whether there exist upper and lower bounds for each or some of these
sequences.

Isocausality may also be helpful in several different fields. For
instance,
since causality is a basic property of our spacetime, some
researchers have attempted to implement it in a theory
at quantum scales and with a discrete ordered set as basic starting
point.
An example is the Causal Set approach, developed by Sorkin et
al. \cite{SORKIN}, which takes the spacetime at small scales to be a
sort of
discrete set in which a binary relation --with the same properties
that the usual causal precedence between points in a spacetime-- has
been
defined. This structure is smoothed out as we go to larger scales,
thus recovering the differentiability of ordinary spacetime and, as is
claimed \cite{SORKIN}, the metric tensor ``up to a conformal factor''.
However, the smoothing procedure 
might well actually lead to one of the many possible
isocausal metrics on a given manifold, or in simpler words, to
an equivalence class of isocausal Lorentzian manifolds, which is a
much larger class.

Yet another application of our construction is to the understanding
of some causal 
properties of quite complicated spacetimes. The idea here is to find
other
{\em simpler} spacetimes which are isocausal to the one under
consideration.
Of course, this is what was achieved by the very popular Penrose
conformal diagrams
\cite{CONF-BOUND,P1,P2} which have been an invaluable tool for
describing the global structure, the causal boundaries, and the
conformal infinity of many important spacetimes,
including the most relevant solutions of
Einstein field equations. Unfortunately, Penrose conformal diagrams
can only
be drawn for spacetimes which are effectively 2-dimensional (such as
spherically
symmetric spacetimes), or for 2-dimensional subspaces of a given
spacetime
(such as the axis in Kerr's geometry). In both situations the
diagrams may not be ``properly conformal'', as one only cares about
the conformal properties of the relevant 2-dimensional part, but not
about the conformal structure of the whole spacetime.
With the new concept of isocausality at hand it is possible to
generalize
Penrose's constructions to more general situations. The basic Penrose
idea was to
embed the original spacetime into a larger one such that the former is
{\em conformally} related to its image in the larger one. The
boundary of
this image in the larger manifold is the conformal boundary, which
may include
both infinity and singularities. As mentioned above, in many cases in
practice one gives up these requirements, due to the impossibility of
finding a truly conformal mapping, and only the conformal
structure of relevant 2-dimensional parts is treated. Our
generalization can be used to surmount these difficulties, to avoid
such unjustified simplifications, to give a rigorous meaning to 
many Penrose diagrams, and to define a new type of generalized diagram.
The idea again is to drop the
conformal condition, which is replaced by the more general causal
equivalence.
Thus, we will put forward the definition of {\em causal extension},
which is
an embedding of the spacetime in a larger one such that the former
is {\em isocausal} to its image in the larger. By these means, we are
also
able to attach a {\em causal boundary} to, and to draw
{\em causal diagrams} of, many spacetimes. As illustrative examples,
we will present
some explicit causal extensions, boundaries and diagrams, and
in particular we will (i) provide a completely rigorous basis and
justification for several traditional
conformal diagrams and (ii) exhibit the causal diagrams for
spacetimes without a known conformal one, such as several classes
of anisotropic non-conformally flat spatially homogeneous models,
including the general case of Kasner spacetimes.

There arise some technical difficulties in the explicit verification
of the isocausality of spacetimes, and some simple criteria are needed
in this sense, as one cannot check {\em all} possible diffeomorphisms
to see if
they are causal relations. Fortunately, the needed mathematical
background
has been recently developed in \cite{S-E,SUP}. In particular, the
null-cone
preserving maps have been thoroughly classified and characterized in
\cite{S-E}
by means of the so-called superenergy tensors \cite{SUP}. Our main
result in this
sense, which will solve most technical difficulties, is that a
diffeomorphism is
a causal relation if and only if the pull-back of the metric tensor
is
a ``future tensor'', that is to say, a tensor with the dominant
property
\cite{S-E,SUP}---i.e.\ satisfying the
``dominant energy condition'' \cite{FF}--. Given that there are very
simple criteria
to ascertain whether a tensor is causal or not \cite{S-E,SUP},
this main technical problem is partly solved. Many specific examples
will be
provided.

The outline of the paper is as follows: in section \ref{sec:CAUSAL
TENSORS} we introduce the notation and review the basics of causal
tensors needed in our work. In section \ref{sec:CR} we define causal
relations, which are the basic objects of this paper, and show their
mathematical properties which naturally leads us to the idea of
causally related and isocausal Lorentzian manifolds. The interplay
between causal and conformal relations as well as what part
of the null cone is preserved under a general causal relation is
thoroughly analyzed in section \ref{sec:CANONICAL}. Section
\ref{sec:APP-CAUSA} deals with the applications of causal relations to
causality theory paying special attention to the global causal
properties shared by isocausal spacetimes and ordering them according
to their causal behaviour.
Finally, the concepts of causal
extension, causal boundary, causal diagrams and causally
asymptotically
equivalent spacetimes are defined with examples in section
\ref{subsect:CB}.
We end up with some conclusions and open questions.

\section{Preliminaries and causal tensors}\label{sec:CAUSAL TENSORS}

Let us introduce the notation and the basic results
to be used throughout this work.
Differentiable manifolds are denoted by
italic capital letters $V , W, U, M, \dots$ and, to our purposes,
all such manifolds (except $M$)  will be connected causally orientable
$n$-dimensional Lorentzian manifolds. Sometimes, the term
``spacetime'' will
also be used for these Lorentzian manifolds.
The metric tensors of $V$ and $W$ will
always be denoted by $\G$ and $\T$, respectively, and
the signature convention is set to $(+, -,\dots ,-)$. $T_x(V)$ and
$T^{*}_x(V)$ will stand respectively for the tangent and cotangent
spaces at $x\in V$, $T(V)$ and $T^{*}(\V)$ denoting the corresponding
tangent and cotangent bundles of $\V$. Similarly
the bundle of $s$-contravariant and $r$-covariant tensors of $V$
is denoted by ${\cal T}^{s}_{r}(V)$. We use boldface letters for
covariant tensors and tensor fields, including exterior forms, and
we put arrows over vectors and vector fields. As is customary, the
same
kernel letter is used for vectors and one-forms related by the
isomorphism
between $T_x(V)$ and $T^{*}_x(V)$ induced by the metric (``raising and
lowering indices''), that is for instance:
$\mathbf{v} =\G (\cdot\, ,\vec{v})$. The closure, interior, exterior,
and boundary of a set $\zeta$ are denoted by $\bar\zeta$, int$\zeta$,
ext$\zeta$, and $\partial\zeta$, respectively. We use ($\subset$)
$\subseteq$
for the (proper) inclusion. Indices $a,b,\dots ,h$
will sometimes be used and they run from $0$ to $n-1$, and
occasionally from
$0$ to $3$. Superscripts $+$ and $-$ will indicate future and past
oriented objects. If $\f$ is a diffeomorphism from $V$ to $W$,
the push-forward and pull-back are written as $\f'$ and $\f^{*}$
respectively.

The hyperbolic structure of the Lorentzian scalar product naturally
splits the elements of $T_x(V)$ into timelike, spacelike, and null,
and as usual we use the term {\it causal} for the vectors (or vector
fields, or curves) which are non-spacelike.
To fix the notation for these Lorentzian cones we set:
\begin{eqnarray*}
\Z^{+}_x&=&\{\X\in T_x(V) : \X\ \mbox{is causal and future directed}
\},\\
\Z_x&=&\Z^{+}_x\cup\Z^{-}_x,\hspace{1cm}
\Z^{+}(V)=\bigcup_{x\in V}\Z^{+}_x
\end{eqnarray*}
with obvious definitions for $\Z^{-}_x$, $\Z^{-}(V)$ and $\Z(V)$. The
null cone $\partial\Z_x$ is the boundary of $\Z_x$ and its elements
are the
null vectors at $x$.
This splitting immediately translates to the causal one-forms and in
fact,
as has been proven in \cite{S-E}, to the whole tensor bundle as
follows
by introducing the following concept.
\begin{defi}
A tensor ${\bf T}\in {\cal T}^{0}_{r}(x)$
has the \underline{dominant property} at $x\in V$ if
$${\bf T}(\vec{u}_{1},\dots ,\vec{u}_{r})\geq 0
\hspace{1cm} \forall \vec{u}_{1},\dots ,\u_{r}\in \Z^+_{x}.$$
\label{def:DP}
\end{defi}
The set of rank-$r$ tensors with the
dominant property at $x\in V$ will be denoted by $\DP^{+}_r|_x$,
whereas $\DP^{-}_r|_x$ is the set of tensors such that $-{\bf T}\in
\DP^{+}_r|_x$. We put $\DP_r|_x\equiv \DP^{+}_r|_x\cup \DP^{-}_r|_x$.
All these definitions extend straightforwardly to
the bundle ${\cal T}^{0}_{r}(V)$ and we define the subsets
$\DP^{+}_r({\cal U})$, $\DP^{-}_r({\cal U})$ and $\DP_{r}({\cal U})$
for an open subset ${\cal U}\subseteq V$ as
\[
\DP^{\pm}_r({\cal U})=\bigcup_{x\in{\cal U}}\DP^{\pm}_r|_x,\, \, \,
\DP_r({\cal U})=\DP^{+}_r({\cal U})\cup\DP^{-}_r({\cal U}).
\]
We can also define tensor fields with the dominant property
over ${\cal U}$ as the sections of
these sets. The same notation will be used for such objects
because the context
will avoid any confusion. Thus, we arrive at the general definition of
causal tensor introduced in \cite{S-E}, see also \cite{benal1,benal2}.
\begin{defi}
The set of {\bf future tensors} on $V$ is given by $\DP^+(V)\equiv
\bigcup_r \DP^+_r(V)$, and analogously for the past. The elements of
$\DP(V)\equiv \DP^+(V)\cup \DP^-(V)$ are called {\bf causal tensors}.
\label{def:causal}
\end{defi}
The simplest example (leaving aside $\r^+\subset \DP^+(V)$) of causal
tensor
fields are the causal 1-forms, which constitute the set $\DP_{1}(V)$
\cite{S-E}. It should be clear that the dual elements of
$\DP_1(V)$ are the vectors of $\Z(V)$. It can be easily seen that
$\DP(V)$ has an algebraic structure of a graded algebra of cones
\cite{S-E,benal1} which generalize the Lorentzian cone $\Z(V)$.

It is important to have some criteria to ascertain whether a given
tensor is in $\DP(V)$. In this paper we will mainly use two of them.
The first one was proven in \cite{S-E} and says that it is enough to
check
the inequality in definition \ref{def:DP} just for null vectors, and
that
the inequality is strict for timelike ones.
\begin{cri}\label{cri:1}
\begin{enumerate}
\item ${\bf T}\in \DP^+_r|_x$ if and only if
${\bf T}(\vec{k}_{1},\dots ,\vec{k}_{r})\geq 0$ for all
$\vec{k}_{1},\dots ,\vec{k}_{r}\in \partial\Z^+_{x}$.
\item  ${\bf T}\in \DP^+_r|_x$ if and only if
${\bf T}(\vec{u}_{1},\dots ,\vec{u}_{r})> 0$ for all
$\vec{u}_{1},\dots ,\vec{u}_{r}\in \Z^+_{x}\setminus \partial\Z^+_x$.
\end{enumerate}
\end{cri}
A simpler and  much more helpful criterion, which will be repeatedly
used
in this paper, is the following (see \cite{SUP} for a proof.)
\begin{cri}
${\bf T}\in \DP^{+}_r(V)$ if and only if
${\bf T}(\vec{e}_{0},\dots ,\vec{e}_{0})\geq \left|
{\bf T}(\vec{e}_{a_{1}},...,\vec{e}_{a_{r}})\right|
\hspace{1mm} \forall a_{1},...,a_{r}\in \left\{0,1,...,n-1\right\}$
in {\em all} orthonormal bases
$\left\{\vec{e}_{0},...,\vec{e}_{n-1}\right\}$
with a future-pointing timelike $\vec{e}_{0}$.
\label{ORTHONORMAL}
\end{cri}
As is clear, this is the reason for the use of the terminology
``dominant'' in definition \ref{def:DP}.
We will also need some partial converses of the above results
given by the next two lemmas. The first is
\begin{lem}
If ${\bf T}(\X,\dots,\X)>0$ for every ${\bf T}\in\DP^{+}_r|_x$ then
$\X\in \Z_x$. Further, if $r$ is odd, then in fact $\X\in\Z^+_x$.
\label{DP}
\end{lem}
\P Suppose on the contrary that $\X$ were a spacelike vector.
Then, there would exist a timelike vector $\vec{u}\in\Z^{+}_x$ such
that
$\G(\vec{u},\X)=0$, hence
$({\bf u}\otimes\dots\otimes{\bf u})(\X,\dots,\X)=0$.
But ${\bf u}\otimes\dots\otimes{\bf u}\in\DP^{+}_r|_x$ in
contradiction.
The second part is immediate by changing $\X$ to $-\X$.\N

The second lemma was not explicitly mentioned but is implicit in
\cite{S-E}.
Here we present
it with its proof. We recall that $\X$ is called an ``eigenvector''
of a 2-covariant tensor ${\bf T}$ if
${\bf T}(\cdot\, ,\X )=\lambda \G (\cdot\, ,\X )$ and
$\lambda$ is then the corresponding eigenvalue.
\begin{lem}
If ${\bf T}\in \DP^{+}_2|_x$ and $\X\in\Z^{+}_x$ then
${\bf T}(\X,\X)=0\ \Longleftrightarrow\X$ is a null eigenvector of
${\bf T}$.
\label{NULL-EIGEN}
\end{lem}
\P Let $\X\in\Z^{+}_x$ and assume ${\bf T}(\X,\X)=0$.
Then since ${\bf T}(\cdot\, ,\X )\in\DP^{+}_1|_x$ \cite{S-E},
we conclude that ${\bf X}$ and ${\bf T}(\cdot\, ,\X )$ must
be proportional which results in $\X$ being a null eigenvector of
${\bf T}$. The converse is straightforward.\N

Several other results from \cite{S-E,SUP} will be introduced along
the paper
when needed.

\section{Causal relations}\label{sec:CR}
Our main concern is to capture the concept of Lorentzian manifolds
compatible
from the {\em causal} point of view. To that end, we put forward our
primary definition
\begin{defi}
Let $\f:V\rightarrow W$ be a global diffeomorphism between
two Lorentzian manifolds. We say that $W$ is
{\bf causally related with $V$ by $\f$}, denoted
$V\prec_{\f}W$, if for every $\X\in\Z^{+}(V)$, $\f' \X\in \Z^{+}(W)$.
$W$ is said to be {\bf causally related with $V$},
denoted simply by $V\prec W$, if there exists $\f$ such that
$V\prec_{\f}W$.
Any diffeomorphism $\f$ such that $V\prec_{\f}W$ is called a
{\bf causal relation}.
\label{PREC}
\end{defi}
In simpler words, what we demand is that the solid Lorentz
cones at all $x\in V$ are mapped by $\f$ to sets contained in the
solid Lorentz cones at $\f (x)\in W$ keeping the time orientation:
$\f '\Z^{+}_{x}\subseteq \Z^{+}_{\f (x)}, \, \, \, \forall x\in V$.

\noindent
{\it Remarks}
\begin{itemize}
\item We must emphasize the fact that the previous definition only
makes sense as a global concept, because every pair of Lorentzian
manifolds $V$ and $W$ are {\it locally causally related},
that is to say, there always exist neighbourhoods $\U_x$, $\U_y$
for every $x\in V$ and $y\in W$ such that $(\U_y,\T)$ is causally
related with $(\U_x,\G)$. This is due to the local equivalence of
the causal properties of any Lorentzian manifold with that of
flat Minkowski spacetime (see \cite{FF,COND} for details),
which can be shown using Riemannian normal
coordinates in appropriate normal neighbourhoods of $x$ and $y$.
\item There exist diffeomorphic Lorentzian manifolds which are
not causally related by any diffeomorphism at all, as we will
show later with explicit examples. This will be written as
$V\not\prec W$
meaning that no diffeomorphism $\f:V\rightarrow W$ is a causal
relation.
\item Observe also that two Lorentzian manifolds can be causally
related
by some diffeomorphisms but not by others, as the next example
illustrates.
\end{itemize}
\subsubsection*{Example 1}\label{ex:1}
\begin{figure}[h]
\epsfxsize=5cm
\hspace{6cm}\epsfbox{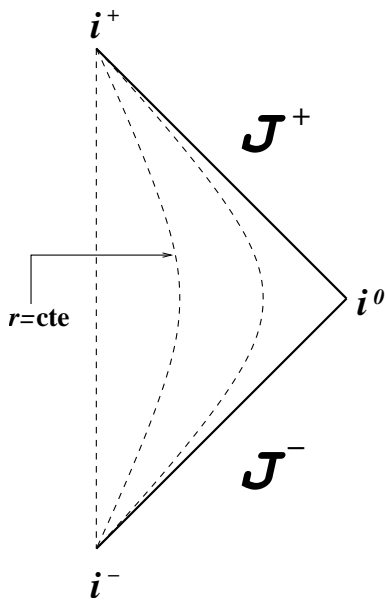}
\caption{Penrose diagram of Minkowski spacetime.
Each point of the diagram represents an $(n-2)$-dimensional sphere of
radius $r$,
except for the vertical line on the left which represents the origin
$r=0$. In all
the figures in this paper, the lines at $45^o$ with respect to the
horizontal
planes are null.}
\label{Minkowski}
\end{figure}

Let $\L$ denote flat Minkowski spacetime. For future reference, we
include
the conformal diagram of $\L$, shown in figure \ref{Minkowski}. Take
standard
Cartesian coordinates
$\{x^{0},x^1,\dots ,x^{n-1}\}$ for $\L$ and consider the
diffeomorphisms
$\f_{b}: \L \longrightarrow \L$ defined by
$$
(x^{0},x^1,\dots ,x^{n-1}) \stackrel{\f_b}{\longrightarrow}
(b\,x^{0},x^1,\dots ,x^{n-1})
$$
for any constant $b\neq 0$. It is easily checked that $\f_{b}$ is a
causal relation for all $b\geq 1$ but not otherwise. Thus
$\L\prec \L$ but, say, $\L\nprec_{\f_{1/2}} \L$. Notice also that for
$b\leq -1$ the diffeomorphisms $\f_{b}$ change the time
orientation of the causal vectors, but still
$\f '\Z_{x}\subseteq \Z_{\f (x)}$, now with
$\f '\Z^{+}_{x}\subseteq \Z^{-}_{\f (x)}$. In such a case, we will
refer
to the diffeomorphism as an {\em anticausal relation}. Obviously any
anticausal relation defines a causal relation by changing the
time orientation of one of the two Lorentzian manifolds. In all the
examples in this paper we will always assume that the explicit time
coordinates increase towards the future.

\subsubsection*{}
Causal relations can be easily characterized by some
equivalent simple conditions.
\begin{prop}
The following statements are equivalent:
\begin{enumerate}
\item $V\prec_{\f} W$.
\item $\f^{*}(\DP^{+}_r(W))\subseteq\DP^{+}_r(V)$ for all $r\in\NAT$.
\item $\f^{*}(\DP^{+}_r(W))\subseteq\DP^{+}_r(V)$ for a given odd
$r\in\NAT$.
\end{enumerate}
\label{CONSERVATION}
\end{prop}
\P

\noindent
$(i)\Rightarrow (ii)$: Let ${\mathbf T}\in \DP_{r}^+(W)$, then
$(\f^{*}{\mathbf T}) (\X_{1},\dots ,\X_{r})=
{\mathbf T}(\f'\X_{1},\dots ,\f'\X_{r})\geq 0$ for all
$\X_{1},\dots ,\X_{r}\in\Z^+(V)$
given that $\f'\X_{1},\dots ,\f'\X_{r}\in \Z^{+}(W)$ by assumption.
Thus $\f^{*}{\mathbf T}\in \DP_{r}^+(V)$.

\noindent
$(ii) \Rightarrow (iii)$: Trivial.

\noindent
$(iii) \Rightarrow (i)$: Fix an odd $r$ and pick up an arbitrary
timelike
$\u\in\Z^{+}(V)$. Then we have:
\[
{\bf T}(\f'\u,\dots,\f'\u)=(\f^{*}{\bf T})(\u,\dots,\u)>0,\
\forall\ {\bf T}\in\DP^{+}_r(W)
\]
since $\f^{*}{\bf T}\in\DP^{+}_r(V)$. Lemma \ref{DP} implies then
$\f'\u\in\Z^{+}(W)$. The result for null $\X\in\Z^{+}(V)$ follows
by continuity.\N

The previous characterizations are natural, but they are not very
useful
as one has to check the property for an entire infinite set of
objects,
as in the original definition \ref{PREC}. Fortunately, a much more
useful
and stronger result can be obtained. Recall that $\T$ is the metric
tensor of $W$.
\begin{theo}
A diffeomorphism $\f:V\rightarrow W$ satisfies
$\f^{*}\T\in\DP^{+}_2(V)$
if and only if $\f$ is either a causal or an anticausal relation.
\label{CAUSAL-CHAR}
\end{theo}
\P By using
\be
\tilde{\G}(\f'\X,\f'\Y)=\f^{*}\tilde{\G}(\X,\Y), \, \,
\forall \X,\Y\in T(V)
\label{pull-back}
\ee
we immediately realize that $V\prec_{\f} W$ implies
$\f^{*}\tilde{\G}\in\DP^{+}_2(V)$, and analogously for the anticausal
case.
Conversely, if $\f^{*}\T\in\DP^{+}_2(V)$ then for every
$\X\in\Z^{+}(V)$
we have that $(\f^{*}\T)(\X,\X)=\T(\f'\X,\f'\X)\geq 0$ hence
$\f'\X\in\Z(W)$. Further, for any other $\vec Y \in\Z^{+}(V)$,
$(\f^{*}\T)(\X,\vec Y)=\T(\f'\X,\f'\vec Y)\geq 0$ so that every pair
of
vectors with the same time orientation are mapped to vectors with the
same time orientation.\N

As we see, it may happen that $\Z^{+}(V)$ is
actually mapped to $\Z^-(W)$, and $\Z^{-}(V)$ to $\Z^+(W)$. As was
explained in the Example 1
one can then always construct
a causal relation by changing, if necessary, the time orientation of
$W$.
Another possibility is to use the following result
\begin{coro}
$V\prec_{\f} W \, \Longleftrightarrow\,
\f^{*}\T\in \DP_{2}^+(V)$ and $\f'\X\in\Z^+(W)$ for at least
one $\X\in\Z^+(V)$.\label{pull-back2}\N
\end{coro}
Leaving this rather trivial time-orientation question aside
(in the end, $\f^{*}\T\in \DP_{2}^+(V)$ always implies that $W$
{\em with one of its time orientations} is causally related with $V$),
let us stress that the theorem \ref{CAUSAL-CHAR} and its corollary are
very powerful, because the condition $\f^{*}\T\in \DP_{2}^+(V)$ is
very easy to check and thereby extremely
valuable in practical problems: first, one only has to work
with one tensor field $\T$, and second, as we saw in the criteria
\ref{cri:1}
and \ref{ORTHONORMAL}, there are several simple ways to check whether
$\f^{*}\T\in\DP_{2}^+(V)$ or not.

Another consequence of the previous theorem is that, for a given
diffeomorphism $\f$, it is enough to demand that $\f'\k$ be causal
just for the null $\k\in \Z^{+}(V)$, as follows from criterion
\ref{cri:1}.
\begin{coro}
$V\prec_{\f} W \, \Longleftrightarrow\,\f'\k\in\Z^{+}(W)$ for all null
$\k\in\Z^{+}(V)$.\label{NULL}\N
\end{coro}

One can be more precise about the causal character of vector fields
and
1-forms when mapped by a causal relation. This will be relevant later
for
the applications to causality theory.
\begin{prop}
If $V\prec_{\f}W$ then
\begin{enumerate}
\item $\X\in\Z^{+}(V)$ is timelike $\Longrightarrow\,\, \f'\X\in
\Z^{+}(W)$
is timelike.
\item $\X\in\Z^{+}(V)$ and $\f^{'}\X\in \Z^{+}(W)$ is null
$\Longrightarrow \X$ is null.
\item ${\mathbf K}\in\DP^+_1(W)$ is timelike $\Longrightarrow\,\,
\f^{*}{\mathbf K}\in \DP^{+}_1(V)$ is timelike.
\item ${\mathbf K}\in\DP^+_1(W)$ and $\f^{*}{\mathbf K}\in
\DP^{+}_1(V)$
is null $\Longrightarrow {\mathbf K}$ is null.
\end{enumerate}
\label{CAUS}
\end{prop}
\P To prove (i) and (ii), theorem \ref{CAUSAL-CHAR} ensures that
$\f^{*}\tilde{\G}\in\DP^{+}_2(V)$ so that for any $\X\in\Z^{+}(V)$
we have, according to equation (\ref{pull-back}), that
$0\leq\f^{*}\tilde{\G}(\X,\X)=\T(\f'\X,\f'\X)$. Using now
criterion \ref{cri:1} to discriminate the strict inequality from the
equality provides the two results. Now, the two other statements
follow
straightforwardly taking into account
$$
(\f^{*}{\mathbf K})(\X)={\mathbf K}(\f^{'}\X),\hspace{5mm}
\forall \X\in T(V)\ \mbox{and}\ {\mathbf K}\in T^{*}(W)
$$
and the fact that ${\mathbf K}$ is null if $\vec K$
is null.\N

Clearly $V\prec V$ for all $V$ by just taking the identity
mapping. Moreover, the next proposition proves that
$\prec$ is transitive too.
\begin{prop} $V\prec W$ and $W\prec U \Longrightarrow  V\prec U$.
\label{ORDER}
\end{prop}
\P There are $\f ,\psi$ such that $V\prec_{\f} W$ and $W\prec_{\psi}
U$ so
that, for any $\X\in \Z^{+}(V)$, $\f'\X\in \Z^{+}(W)$ and
$\psi'[\f'\X]\in\Z^{+}(U)$. Hence
$(\psi\circ\f)^{'}\X\in\Z^{+}(U)$ from where $V\prec U$.\N

It follows that the binary relation $\prec$ is a preorder for the
class
of all diffeomorphic Lorentzian manifolds.
This is not a partial order as $V\prec W$ and
$W\prec V$ do not imply that $V=W$.
This allows us to put forward the following
\begin{defi}
Two Lorentzian manifolds $V$ and $W$ are called causally equivalent,
or
in short {\bf isocausal}, if $V\prec W$ and $W\prec V$. This will be
denoted by $V\sim W$.\label{EQUIV}
\end{defi}
The fact that $V\sim W$ does not imply that $V$ is conformally
related to $W$, as we will prove explicitly in the next section and with
examples. The point here is  that $V\prec_{\f} W$ and $W\prec_{\psi} V$ can
perfectly happen with $\psi \neq \f^{-1}$. Nevertheless, if $V\sim W$ 
both spacetimes are {\em mutually} casually compatible and we will show that some
 global causal properties are shared by $V$ and $W$.
\subsubsection*{Example 2}\label{ex:2}
\begin{figure}[h]
\epsfxsize=6cm
\hspace{6cm}\epsfbox{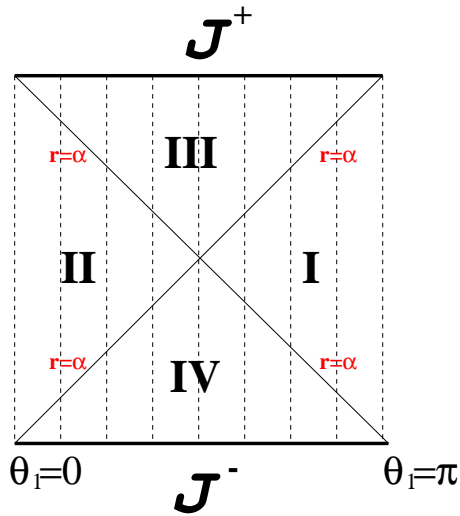}
\caption{\label{QUART} Penrose diagram of de Sitter spacetime. Each
point
represents a $(n-2)$-sphere except for the two vertical lines at
$\theta_1=0,\pi$.
With respect to the chosen time coordinate we show the static regions
I and II
and the non-static ones III and IV. Observe that de Sitter spacetime is
homogeneous and therefore all its points are equivalent. This would
be clearer if we replace each horizontal line by a
circle ---representing the $S^{n-1}$---in the same manner as
in Fig.\ref{EINSTEIN}. Then the conformal diagram will be given only
by the surface of the resulting truncated cylinder. In that case,
all points will be equivalent and we can clearly choose
the origin of coordinates at any vertical line. With respect to this
choice, the coordinates of equation (\ref{ONE-FOURTH}) cover any
one of the static regions.
It is noteworthy to point out the resemblance of the Penrose diagram of
each static region with the diagram of Minkowski spacetime.
Similarly, the
union of any one of the static regions with one of the non-static
ones provides the Penrose diagram for
the so-called steady state model of the Universe, up to time
orientation.}
\end{figure}

Let us denote by $\E$ the Einstein static universe and by $\dS$ the de
Sitter spacetime, both in general dimension $n$, whose base
differential manifold is $\r\times S^{n-1}$ and hence they are
diffeomorphic.  The corresponding line-elements are, with
$a,\alpha$=constants: \bea (\dS,\G) :& & ds^{2}=dt^{2}-
\alpha^{2}\cosh^{2}(t/\alpha)d\O^{2}_{n-1},\,\,\, \alpha>0 \nonumber\\
(\E,\T) :& & d\tilde{s}^{2}=d\bar{t}^{2}-a^2d\bar{\O}^{2}_{n-1}\
\nonumber \eea where $d\O^2_{n-1}$ (and its barred version
$d\bar{\O}^2_{n-1}$) is the canonical round metric in the
$(n-1)$-sphere $S^{n-1}$, given by
$$
d\O^2_{n-1}=d\z_1^2+\sin^2\z_1(d\z^{2}_2+\sin^2\z_2d\z^{2}_3+\dots +
\sin^2\z_2\cdots\sin^2\z_{n-2}d\z^2_{n-1})
$$
with the angles running in the intervals $0<\z_{k}<\pi$
for $k=1,\dots ,n-2$, and $0<\z_{n-1}<2\pi$. For future reference, we
include
the conformal diagram of $\dS$, shown in Figure \ref{QUART}.

Define the diffeomorphisms
$\f_b:\dS\rightarrow \E$ by
$(\bar{t},\bar{\z_{i}})=(bt,\z_{i}),\ b>0$. Then
$$
\f^{*}_b\T =\left(b^{2}-\fr{a^2}{\alpha^{2}\cosh^{2}(t/\alpha)}\right)
dt\otimes dt +\fr{a^2}{\alpha^{2}\cosh^{2}(t/\alpha)}\G
$$
so that by using any of the criteria \ref{cri:1} or \ref{ORTHONORMAL}
one can easily check that $\f^{*}_b\T\in\DP^{+}_{2}(\dS )$ if
$b^{2}\geq a^2/\alpha^{2}$. The corollary \ref{pull-back2} immediately
implies then that $\f_b$ are causal relations for these values of
$b$, so that
$\dS\prec \E$. A natural question arises: is $\E\prec\dS$ and thus
$\dS\sim\E$?
To answer this question one can try to build an explicit causal
relation from
$\E$ to $\dS$, but one readily realizes that there are no such simple
diffeomorphisms. Of course, at this stage one is unsure whether there
may be
other, yet untried, diffeomorphisms which are the sought causal
relations. But
the problem is the impossibility to check {\em all} the
diffeomorphisms explicitly. Nevertheless, we will prove in section
\ref{sec:APP-CAUSA} that one can find results and criteria allowing to
avoid this
problem completely, and providing very simple ways to prove, or
disprove, the
causal relationship between given spacetimes. Thus, we will answer the
question of whether or not $\dS\sim\E$ in the Example 6 of section
\ref{sec:APP-CAUSA}.

\subsubsection*{Example 3}\label{STATIC-DS}
Take again ordinary $n$-dimensional flat spacetime $\L$ but
now
in spherical coordinates $\{T,R,\z_1,\dots,\z_{n-2}\}$ so that the
line element
reads
\be
(\L,\G): ds^2=dT^2-dR^2-R^2d\O^2_{n-2} \label{flatsph}
\ee
with $-\infty < T< \infty$ and $0<R<\infty$.
The second spacetime will be one of the static regions of de Sitter
spacetime, denoted here by $\fr{1}{4}\dS$, given by the line element
\be
\left(\mbox{$\fr{1}{4}$}\dS,\T\right): d\tilde{s}^{2}=
\left(1-\fr{r^{2}}{\alpha^{2}}\right)dt^{2}-
\left(1-\fr{r^{2}}{\alpha^{2}}\right)^{-1}dr^2-r^{2}d\bar{\O}^2_{n-2},
\label{ONE-FOURTH}
\ee
where the non-angular coordinate ranges are $-\infty<t<\infty,\
0<r<\alpha$
(see figure \ref{QUART}).
We are going to show that these spacetimes are causally equivalent.
To that end, consider the diffeomorphisms
$\f_{\beta}:\L\rightarrow \fr{1}{4}\dS$ and
$\p_f:\fr{1}{4}\dS\rightarrow \L$ defined by
\begin{eqnarray*}
(T,R,\z_k)\stackrel{\f_\beta}\longrightarrow (T,\fr{\alpha\beta R}{1+\beta R},\z_k)\\
(t,r,\bar{\z_k})\stackrel{\p_{f}}\longrightarrow (t,f(r),\bar{\z_k})
\end{eqnarray*}
where $\beta$ is a positive constant and $f(r)$ a function to be
determined.
By writing down $\f^{*}_{\beta}\T$ and $\p^{*}_f\G$ in
appropriate orthonormal cobases we obtain their eigenvalues with
respect
to $\G$ and $\T$, given respectively by
(we shall always write the ``timelike'' eigenvalue first)
\begin{eqnarray*}
\left\{\fr{1+2\beta R}{(1+\beta R)^{2}},
\fr{\alpha^2\beta^2}{(1+\beta R)^{2}(1+2\beta R)},
\fr{\alpha^2\beta^2}{(1+\beta R)^{2}},\dots ,
\fr{\alpha^2\beta^2}{(1+\beta R)^{2}}\right\},\\
\left\{\left(1-\fr{r^{2}}{\alpha^{2}}\right)^{-1},
\left(1-\fr{r^{2}}{\alpha^{2}}\right)f'^{2},
\fr{f^2}{r^2},\dots ,\fr{f^2}{r^2}\right\}.
\end{eqnarray*}
By using now criterion \ref{ORTHONORMAL} we can write down the
conditions for
$\f^{*}_{\beta}\T$ and $\p^{*}_f\G$ to be in $\DP^{+}(\L)$ and
$\DP^{+}(\fr{1}{4}\dS)$ respectively:
\bea
\f^{*}_{\beta}\T\in\DP^{+}_{2}(\L)\Longleftrightarrow
\beta^{2}\leq\fr{1}{\alpha^2}
\label{nova1}\\
\p^{*}_{f}\G\in\DP^{+}_2(\mbox{$\fr{1}{4}$}\dS)\Longleftrightarrow
\left(1-\fr{r^2}{\alpha^{2}}\right)^{-2}\geq f'^{2},\
\left(1-\fr{r^2}{\alpha^2}\right)^{-1}\geq\fr{f^2}{r^2}.\label{nova2}
\eea
Condition (\ref{nova1}) is clearly satisfied choosing the values of
$\beta$,
while condition (\ref{nova2}) is easily seen to be fulfilled for
suitable
choices of $f(r)$. One such choice is, for instance,
$f(r)= -br\log\left|1-\fr{r^2}{\alpha^2}\right|$, for adequate values of
the
constant $b$. Finally, corollary \ref{pull-back2} ensures then that
$\fr{1}{4}\dS\sim\L$.

This example illustrates how two Lorentzian manifolds with different
global and metric properties can be isocausal. Notice that
$\fr{1}{4}\dS$ is
geodesically incomplete while $\L$ is b-complete (see
\cite{BEE,FF,COND}), and
nevertheless they are isocausal. As we see from figures
\ref{Minkowski} and
\ref{QUART}, the Penrose diagrams of both spacetimes have a similar
``shape''.
We will provide further examples, starting with the next Example 4,
showing that this happens in general for
causally equivalent spacetimes if their Penrose conformal diagrams are
defined. Thereby, the causal equivalence can provide an adequate
generalization, for cases in which the Penrose diagrams cannot be drawn,
of these very useful drawings/representations of spacetimes. We will
present several examples in this paper.

\subsubsection*{Example 4}\label{PRESSURE}
\begin{figure}[h]
\epsfxsize=\textwidth
\epsfbox{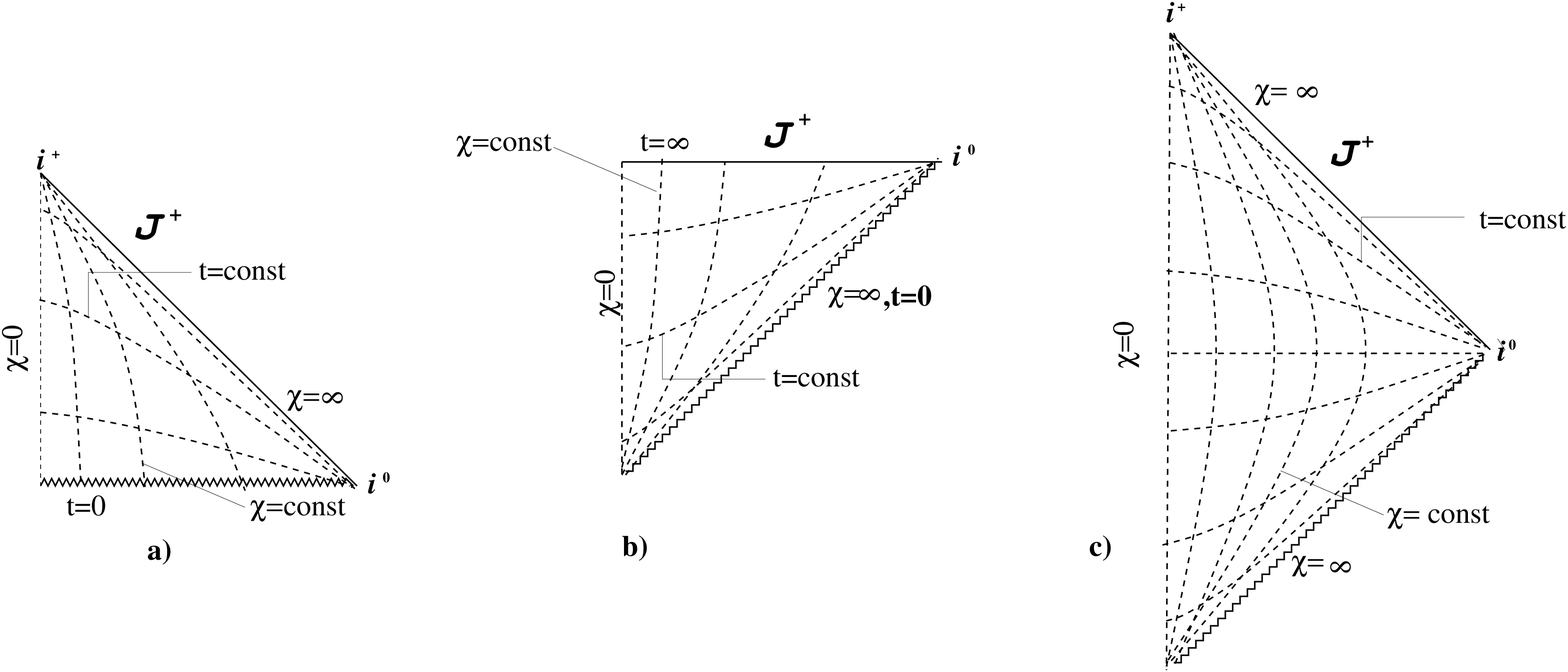}
\caption{\label{STEADY} Penrose diagrams of $\RW_0\{\g\}$ spacetimes
for
$\g\in (-\infty,-1)\cup (\fr{3-n}{n-1},\infty)$ (case a), $-1<\g<
\fr{3-n}{n-1}$ (case b),
and $\g=\fr{3-n}{n-1}$ (case c).
Notice the similar shape of these diagrams with that of Minkowski
spacetime (case c) and of the steady state spacetime up to time
orientation (cases a and b).}
\end{figure}

Let us consider the $n$-dimensional Robertson-Walker spacetimes $\RW_k$
\cite{FF,COND} for the case
of flat spatial sections ($k=0$) and such that the equation of state
for the
cosmological perfect fluid is $p=\g\rho$ where $p$ is the isotropic
pressure,
$\rho$ is the energy density and $\g$ is a constant.
Solving the Einstein equations under these hypotheses the scale
factor
takes the form $a(t)=Ct^{\fr{2}{(n-1)(1+\g)}}$ where $C$ is a constant and
$\g\neq -1$, see e.g. \cite{COND} for $n=4$. Hence the line-element is given by
\be
(\RW_{0}\{\g\},\T):\
d\tilde{s}^2=dt^2-C^2t^{\fr{4}{(n-1)(1+\g)}}(d\x^2+\x^2d\bar{\O}^2_{n-2}),\,\,
0<t,\chi <\infty .
\label{FRLW}
\ee
The Penrose diagrams of these spacetimes are shown in figure
\ref{STEADY} for every value of $\g$.

The exceptional case $\g =-1$ is in fact the part of de Sitter universe
$\dS$ usually called the ``steady state'' model in General Relativity
\cite{FF} and shown in figure \ref{QUART},
which we will denote here by $\fr{1}{2}\dS =\RW_0\{-1\}$.
Its line-element reads
\be
\hspace{-1.5cm}\left(\mbox{$\fr{1}{2}$}\dS_{\pm},\G\right) :\,\,
ds^2=dT^2-e^{\pm 2 T/\alpha}(dR^2+R^2d\Omega^2_{n-2}),\hspace{3mm}
 -\infty<T<\infty,\ 0<R<\infty \, \label{sstate}
\ee
where now the coordinates cover only the regions II and IV
(for the minus sign), or II and III (for the plus sign),
of the full de Sitter spacetime shown in figure \ref{QUART}.

As conjectured in the previous example, spacetimes with equal-shaped
conformal
diagrams will be isocausal. Therefore, from figures \ref{QUART} and
\ref{STEADY} we guess that $\RW_0\{\g\in(-1,\fr{3-n}{n-1})\}$ will be
isocausal to
$\fr{1}{2}\dS_+$ (with the plus sign), while
$\RW_0\{\g\notin [-1,\fr{3-n}{n-1}]\}$ will be isocausal to
$\fr{1}{2}\dS_-$ . The remaining case $\g =\fr{3-n}{n-1}$ has a diagram which
is in fact similar to that of $\fr{1}{4}\dS$, i.e. the static region II of
de Sitter spacetime, see figure \ref{QUART}, which we already know to be
isocausal to flat
spacetime $\L$. Thus, $\RW_0\{\fr{3-n}{n-1}\}$ will be isocausal to $\L$.
We are now going to prove that all these conjectures are actually
true.

To that end, and without loss of generality, we put $C=1$ in
(\ref{FRLW})
and $\alpha =1$ in (\ref{sstate}). The candidate diffeomorphisms
$\f_{b,c} : \L \longrightarrow \RW_0\{\g\}$,
or $\f_{b,c} : \fr{1}{2}\dS_{\pm} \longrightarrow \RW_0\{\g\}$,
will be defined by
$$
(T,R,\z_i)\stackrel{\f_{b,c}}\longrightarrow (be^{cT},R,\z_i)
$$
for some constants $b$ and $c>0$. Thus we respectively get
$$
\f^{*}_{b,c}\T =\left(b^2c^{2}e^{2cT}-
K\exp\left(\fr{4Tc}{(n-1)(1+\g)}\right)\right)
dT\otimes dT +K\exp\left(\fr{4Tc}{(n-1)(1+\g)}\right)\G |_{\L}
$$
$$
\hspace{-.5cm}\f^{*}_{b,c}\T =\left(b^2c^{2}e^{2cT}\!\!\!-K
\exp\left(\fr{4cT}{(n-1)(1\!+\!\g)}\!\mp\!\! 2T\right)\right)dT\otimes dT +
K\exp\left(\fr{4cT}{(n-1)(1\!+\!\g)}\!\mp\!\! 2T\right)\G |_{\fr{1}{2}\dS_{\pm}}
$$
where $K=b^{\fr{4}{(n-1)(1+\g)}}$. Therefore, using criteria \ref{cri:1} or
\ref{ORTHONORMAL}, $\f^{*}\T\in\DP^{+}_2(\L)$ in the first case if
and only if
$b^2c^2e^{2cT}\geq K\exp\left(\fr{4c}{(n-1)(1+\g)}T\right)$
holds for every value of $T$, and this happens only if
$\g=\fr{3-n}{n-1}$ and $c^2\geq 1$.
Choosing $c=b=1$ (say), from corollary \ref{pull-back2} we have
$\L\prec_{\f_{1,1}}\RW_0\{\fr{3-n}{n-1}\}$.
Notice that in this case we have then $\f^{*}_{1,1}\T\propto \G$,
i.e.,
$\f_{1,1}$ is a conformal relation, see the next section. In this case
$\f^{-1}_{1,1}$ is also a causal relation, as can be easily checked,
and therefore $\L\sim\RW_0\{\fr{3-n}{n-1}\}$.

Similarly,
$\f^{*}_{b,c}\T\in\DP^{+}_2(\fr{1}{2}\dS_{\pm})$ in the second case if and
only if, $\forall T\in (-\infty,\infty)$,
$b^2c^2\geq K\exp\left[\left(\fr{4c}{(n-1)(1+\g)}\mp 2 -2c\right)T\right]$
which holds for appropriate values of $b$ whenever
$c=\mp\fr{(n-1)(1+\g)}{n-3+(n-1)\g}$.
As $c$ must be a positive constant so that
the causal orientations are made consistent, we must choose
the plus sign for $\g\in(-1,\fr{3-n}{n-1})$ and the minus sign for $\g\notin
[-1,\fr{3-n}{n-1}]$.
If we choose $b$ such that $b^2c^2=K$ then the inverse diffeomorphism
is also a causal relation and we have
$\fr{1}{2}\dS_-\sim \RW_0\{\g\notin [-1,\fr{3-n}{n-1}]\}$, and
$\fr{1}{2}\dS_+\sim \RW_0\{\g\in(-1,\fr{3-n}{n-1})\}$.

\section{Canonical null directions of causal relations. Conformal
relations}
\label{sec:CANONICAL}
As was already pointed out, if $\f$ is a causal relation between $V$
and $W$,
then the Lorentzian cone of a point in $V$ is mapped by means of $\f$
within
the Lorentzian cone of the image point of $W$. Nevertheless, as we
have seen
in the previous example, there are cases in which the causal
relations are
conformal and then the null cones (that is, the boundaries of the
Lorentz cones)
are preserved. In general, a part of the initial null cone may or may
not
remain on the final null cone by the application of $\f$, but those
parts
which do remain can be identified easily by means of the next result.
\begin{prop}
Let $V\prec_{\f} W$ and $\X\in \Z^{+}_x$. Then
$\f'\X\in\partial\Z^+_{\f (x)}$
if and only if $\X$ is a null eigenvector of $\f^{*}\tilde{\G}|_{x}$.
\label{CONE}
\end{prop}
\P Let $\X$ be an element of $\Z^{+}_x$ and suppose $\f'\X$ is null
at $\f(x)$.
Then according to proposition \ref{CAUS} $\X$ is also null at $x$.
On the other hand we have
\[
0=\tilde{\G}|_{\f(x)}(\f^{'}\X,\f^{'}\X)=\f^{*}\tilde{\G}|_{x}(\X,\X)
\]
and since $\f^{*}\T|_{x}\in\DP^{+}_2(x)$, lemma \ref{NULL-EIGEN}
implies that $\X$ is a null eigenvector of $\f^{*}\T$ at $x$. The
converse
is trivial.\N

The existence of null vectors which remain null under the application
of a
causal relation motivates the next definition.
\begin{defi}
If the relation $V\prec_{\f} W$ holds and $\f^{*}\T$ possesses $m$
independent
null eigenvectors $\forall x\in V$, these are called the
{\bf canonical null directions} of $\prec_{\f}$.\label{CURVE}
\end{defi}

\noindent
{\it Remarks}
\begin{itemize}
\item The importance of proposition \ref{CONE} and definition
\ref{CURVE} lies
on the recently proved fact that the null eigenvectors of any tensor
in
$\DP^{+}_2$ thoroughly classify it by means of its canonical
decomposition
found in \cite{S-E}. The relevant result here is Theorem 4.1 of
\cite{S-E},
which can be summarized as
\begin{theo}
Every ${\bf T}\in\DP^{+}_2(V)$ can be written canonically as the sum
${\bf T}=\sum^{n}_{r=1}{\bf S}\{{\bf \O}_{[r]}\}$ of rank-2
``super-energy tensors'' ${\bf S}\{{\bf \O}_{[r]}\}\in \DP^+_2(V)$ of
simple $r$-forms ${\bf \O}_{[r]}$. Furthermore, the decomposition is
characterized by the null eigenvectors of ${\bf T}$ as follows:
if ${\bf T}$ has $m$ linearly
independent null eigenvectors $\k_1, \dots, \k_m$ then the sum starts
at
$r=m$ and ${\bf \O}_{[m]}={\bf k}_1\wedge\dots\wedge {\bf k}_m$; if
${\bf T}$
has no null eigenvector then the sum starts at $r=1$ and
$\vec{\O}_{[1]}$ is
the timelike eigenvector of ${\bf T}$.\label{CLASS}\N
\end{theo}
For the sake of completeness, let us recall that the
{\it super-energy tensor} of an arbitrary $r$-form ${\bf \Lambda}$ is
given by the
formula \cite{SUP}:
\be
S_{ab}\{{\bf \Lambda}\}=\fr{(-1)^{r-1}}{(r-1)!}
\left[\Lambda_{aa_2\dots a_r}\Lambda_{b}^{\ a_2\dots a_r}-
\fr{1}{2r}\Lambda_{a_1\dots a_n}\Lambda^{a_1\dots a_n}\gg_{ab}\right]
\label{decompos}
\ee
and in general they satisfy ${\bf S}\{{\bf \Lambda}\}\in \DP^+_2(V)$
and
$S_{ab}\{{\bf \Lambda}\}=S_{ba}\{{\bf \Lambda}\}$. If ${\bf \Lambda}$
is a
simple $r$-form then ${\bf S}\{{\bf \Lambda}\}$ is proportional to an
{\it involutory Lorentz transformation} because
$S_{ac}S^{c}_{\ b}\propto \gg_{ab}$. We deduce from this theorem and
equation
(\ref{decompos}) that any tensor of $\DP^{+}_{2}(V)$ possessing $\n$
independent
null eigenvectors is the metric tensor up to a positive factor.
See \cite{S-E} for further details.
\item
Therefore, if $V\prec_{\f}W$ then $\f^*\T$ (which is in $\DP^+_2(V)$
by
theorem \ref{CAUSAL-CHAR}) admits always a decomposition of the type
shown
in theorem \ref{CLASS}, and the number of its canonical null
directions,
if there are any, is given by the number $r$ where that sum starts.
\end{itemize}

With the aid of the previous remarks we get an important theorem
which characterizes the conformal relations among the set of all
causal
relations between Lorentzian manifolds.
\begin{theo}
For a diffeomorphism $\f : V \longrightarrow W$ the following
properties
are equivalent, characterizing the conformal relations:
\begin{enumerate}
\item $\f$ is a causal (or anticausal) relation with $n$ canonical
null directions.
\item $\f^* \T =\lambda \G$, $\lambda >0$.
\item $(\f^{-1})^* \G =\mu \T$, $\mu>0$.
\item $\f$ and $\f^{-1}$ are both causal (or both anticausal)
relations.
\end{enumerate}
\label{CONF}
\end{theo}
\P

\noindent
$(i) \Rightarrow (ii)$ If $\f$ is a causal relation with $n$
independent
canonical null directions, then $\f^{*}\T\in\DP^+_2(V)$ has $n$
independent
null eigenvectors which is only possible,
according to theorem \ref{CLASS} and its remarks, if
$\f^{*}\T=\lambda \G$ for some positive function $\lambda$ defined on
$V$.

\noindent
$(ii) \Leftrightarrow (iii)$ If $\f^{*}\T=\lambda \G$, then
$\T=(\f\circ\f^{-1})^*\T=(\f^{-1})^*\f^*\T=(\f^{-1})^*(\lambda \G)$.
The converse is similar.

\noindent
$(iii) \Rightarrow (iv)$ Theorem \ref{CAUSAL-CHAR} together with
$(ii)$ and $(iii)$ imply $(iv)$ immediately.

\noindent
$(iv) \Rightarrow (i)$ If $(iv)$ holds, we can establish the
following assertion by application of proposition \ref{CAUS} to
$\f^{-1}$
$$
(\f^{-1})^{'}\Y\in\Z^{+}(V)\
\mbox{is null and $\Y\in\Z^{+}(W)$}\Longrightarrow\Y\ \mbox{is null}.
$$
Now, let $\X\in\Z^{+}(V)$ be null and consider the unique $\Y\in T(W)$
such that $\X=(\f^{-1})^{'}\Y$.
Then $\Y=\f^{'}\X$ and $\Y\in\Z^{+}(W)$ because
$\f$ is a causal relation (the anticausal case is similar).
According to the assertion above $\Y$ must then be null and we
conclude
that every null $\X\in\Z^{+}(V)$ is push-forwarded to a null vector
of $\Z^{+}(W)$. Thus, proposition \ref{CONE} implies in fact that all
null
vectors are eigenvectors of $\f^*\G$.\N

This theorem fully characterizes the (time-preserving) conformal
relations as those diffeomorphisms
mapping null future-directed vectors onto null future-directed
vectors.
It is worth remarking here that there are a number of results
characterizing
conformal relations as the {\em homeomorphisms} preserving the null
geodesics, see \cite{CF1}, \cite{CF3}.

Observe that theorem \ref{CONF} implies that $V\prec_{\f}W$ and
$W\prec_{\f^{-1}}V$ hold if and only if $\f$ is a conformal relation.
Thus, as was naturally expected, if $\f:V\rightarrow W$ is a conformal
relation, then $V\sim W$. However, the converse does not hold in
general,
and there are isocausal spacetimes which are not conformally related.
This
happens when $V\prec_{\f}W$ and $W\prec_{\psi}V$, but
$W\nprec_{\f^{-1}}V$.
In consequence, the causal equivalence is a generalization of the
conformal relation between Lorentzian manifolds. 

A door open by theorem \ref{CONF} is the question of whether one can
consistently define the concept of  ``partly conformal'' Lorentzian
manifolds among those which are isocausal. The idea here is to
explore the
possibility of having conformally related subspaces without the full
manifolds
being conformal. This idea can be made precise as follows
\begin{defi}
If $V\sim W$, we shall say that $V$ and $W$ are $\fr{m}{n}$-{\bf
conformally
related} if there are causal relations $V\prec_{\f}W$ and
$W\prec_{\psi}V$ with
$m$ corresponding canonical null directions.
\end{defi}
{\it Remarks}
\begin{itemize}
\item By ``corresponding'' canonical null directions we mean that $m$
null
eigenvectors of $\f^*\T$ are mapped by $\f$ to $m$ null eigenvectors
of
$\psi^* \G$, and vice versa.
\item In general, two isocausal spacetimes are not conformally
related at all.
However, if they are $\fr{m}{n}$-conformally related, then they are
$\fr{s}{n}$-conformally related for all natural numbers $s\leq m$.
Thus,
the sensible thing to do is to speak about $\fr{m}{n}$-conformal
relations
only for the maximum value of $m$.
\item Obviously, the $\fr{n}{n}$-conformal relation is just the
conformal
relation. We know that two locally  conformal spacetimes are
characterized by the
preservation of the so-called conformal (Weyl) curvature tensor
\cite{FF,Eis}.
The generalization to the case of partial conformal relations is
under current
investigation \cite{letter}.
\end{itemize}

\subsubsection*{Example 5}\label{ppwaves}
Consider the general form of the line-element for the so-called
``pp-waves'',
see e.g. \cite{KRA,PW}, given in general dimension $n$ by
\be
\hspace{-0.5cm}ds^2=2dudv-\sum_{k=2}^{n-1}(dx^k)^2 +2H_1(u,x^k)du^2,\
-\infty<u,v,x^k<\infty .\label{ppw}
\ee
The base manifold of these spacetimes is $\r^n$ and we will denote
them by
$\ppW (H_1)$. Let $\ppW (H_2)$ be another pp-wave spacetime
with a different function $H_2$
and coordinates $\{\bar{u},\bar{v},\bar{x}^k\}$.
To compare them causally, take the diffeomorphisms
$\f_{f}:(\ppW (H_1),\G)\rightarrow (\ppW (H_2),\T)$ and
$\psi:(\ppW (H_2),\T)\rightarrow (\ppW (H_1),\G)$ as follows:
$$
(u,v,x^k)\stackrel{\f_f}\longrightarrow (u,v+f(u),x^k), \hspace{1cm}
(\bar{u},\bar{v},\bar{x}^k)\stackrel{\psi}\longrightarrow
(\bar{u},\bar{v},\bar{x}^k)
$$
so that a simple calculation gives
$$
\f_f^{*}\T=\G+2\left(H_2 -H_1 +f'(u)\right)\, {\bf k}\otimes{\bf k} ,
\hspace{1cm}
\psi^{*}\G=\T+2(H_1 -H_2 )\, \bar{{\bf k}}\otimes\bar{{\bf k}}
$$
where ${\bf k}=du$ is a future-directed
null 1-form in $\ppW(H_1)$ and the same for ${\bar{\bf k}}=d\bar{u}$.
It is then clear, by using criterion \ref{cri:1}, that
$\f_f^{*}\T\in\DP^+(\ppW (H_1))$ if and only if $H_2-H_1+f'\geq 0$,
and
that $\psi^{*}\G\in\DP^+(\ppW (H_2))$ iff $H_1-H_2 \geq 0$. Hence, due
to corollary \ref{pull-back2}, the two pp-wave spacetimes will be
isocausal if, for instance,
$$
0\leq H_1 -H_2 \leq f'(u) .
$$
There are many possibilities to comply with such conditions, one
simple example
is $H_1 -H_2=\sin^2 F$ and $f(u)=\sinh u$, where $F(u,x^k)$ is an
arbitrary
function.
In this case, they are in fact $\fr{1}{n}$-conformally related,
because the
null vectors $\vec k$ and $\vec{\bar{k}}$ are corresponding canonical
null
directions for those diffeomorphisms, as can be easily checked: they
are null
eigenvectors of $\f_{\sinh u}^{*}\T$ and $\psi^{*}\G$, respectively.

We may note in passing that this can be used to provide
an explicit example of a pair of isocausal spacetimes not conformally 
related (not even locally). For if we take $H_{2}=0$ so that $\ppW (0)=\L$ 
is Minkowski spacetime, the condition for isocausality
becomes $0\leq H_1\leq f^{'}(u)$ and it is very easy to choose $H_1$
in such a way that $\ppW (H_1)$ is not locally conformally flat.

\section{Applications to causality theory}
\label{sec:APP-CAUSA}
In this section we study how the causal properties
of two Lorentzian manifolds $V$ and $W$ are related when $V\prec W$.
For that purpose, let us recall the basic sets used in causality
theory \cite{BEE,FF,COND,Wald}.
If $p,q\in V$, $p<q$ means that there exists a
continuous\footnote{Continuous causal curves are well--defined, see
e.g. \cite{BEE,FF,COND,Wald}.} future-directed
causal curve from $p$ to $q$, and similarly for $p<\!\!<q$ if the
curve can
be timelike. Then the chronological and causal futures of any point
$p$ are
defined respectively by \cite{FF}
\[
I^{+}(p)=\{x\in V : p<\!\!<x \},\ \ J^{+}(p)=\{x\in V : p<x \}
\]
and dually for the past. These definitions are translated in an
obvious way
to arbitrary sets $\zeta\subset V$ and so we write $I^{\pm}(\zeta)$
and
$J^{\pm}(\zeta)$. A set $\zeta$ is called a future set if
$I^{+}(\zeta)\subseteq \zeta$.
For example $I^{+}(\zeta)$ is a future set for any
$\zeta$. A set $\zeta$ is achronal if $\zeta\cap
I^{+}(\zeta)=\emptyset$, and
acausal if there are no points $p,q\in\zeta$ such that $p<q$ (this
implies
that $\zeta\cap J^{+}(\zeta)=\zeta$, but is not equivalent to that in
general.)
The boundary of a future set is always achronal and is called an
achronal boundary\addtocounter{footnote}{+3}\footnote{Sometimes
these sets are referred to as
{\em proper} achronal boundaries \cite{COND}
to distinguish them from achronal sets which are the
boundary of non-future sets \cite{PENROSE,COND}.}. Due to the
connectedness
of the manifold, $V$ can be disjointly decomposed as
$V=\B^{+}\cup\B\cup\B^{-}$ where $\B^{+}$ is any open future set,
$\B$ its achronal
boundary, and $\B^{-}=\mbox{ext}\B^{+}$ is a past set. Of course $\B$
is also
the achronal boundary for $\B^-$. Finally we must also recall the
definitions
of the future and past Cauchy developments. Let $\g^{\pm}_{p}$ be a
future
(past) causal curve passing through $p$ and denote by
$\Gamma^{\pm}_{p}$ the set
of all such {\em endless} curves. The future Cauchy development of
$\zeta$ is
defined as follows
\[
D^{+}(\zeta)=\{x\in V :\,
\g^{-}_{x}\cap\zeta\neq \emptyset\ \forall
\g^{-}_{x}\in\Gamma^{-}_{x}\}
\]
and similarly for $D^{-}(\zeta)$. The Cauchy development of $\zeta$
is then
$D(\zeta)=D^{+}(\zeta)\cup D^{-}(\zeta)$.
All the above
concepts are standard, well studied and defined in many references,
see for
instance \cite{BEE,FF,Wald,COND}.

\subsection{Causality sets and causal relations}
With all the nomenclature now at hand, we can prove several results
giving the
behaviour of the causality sets under the application of a
causal relation between Lorentzian manifolds.
\begin{prop}
$V\prec_{\f} W$ if and only if  every continuous future-directed timelike
(causal)
curve in $V$ is mapped by $\f$ to a continuous future-directed timelike (causal)
curve in $W$.
\label{curves}
\end{prop}
\P If every future-directed timelike curve $\g\subset V$ is mapped by $\f$
to a future-directed timelike curve $\f(\g)\subset W$, then by
choosing the $\g$'s to be $C^1$
every future-directed timelike tangent vector is mapped to a
future-directed timelike vector. As a consequence if $\k\in T(V)$ is null and
future-directed
then $\f'\k$ must be causal and future-directed
(to see this just construct a sequence of
future-directed timelike vectors converging to $\k$.)
Conversely, take any continuous future-directed $\g\subset V$. It is
known that $\g$ must be differentiable almost everywhere \cite{PENROSE}
so that from proposition \ref{CAUS} $\f(\g)$ is
continuous and future-directed almost everywhere. Finally, if $\g$ is
not differentiable at $p\in \g$, then there is a normal
neighbourhood $\U_{p}$ of $p$ such that, for every
$q,r\in\g\cap\U_{p}$ there is a future-directed differentiable arc from
$q$ to $r$. As $V\prec_{\f} W$ this arc is mapped to another
differentiable arc which is future-directed, so that $\f(\U_{p})$ is
a normal neighbourhood of $\f(p)$ with the required property such
that $\f(\g)$ is also continuous and future-directed at $\f(p)$.\N
\begin{prop}
If $V\prec_{\f} W$ then $\f(I^{\pm}(\zeta))\subseteq
I^{\pm}(\f(\zeta))$ and $\f(J^{\pm}(\zeta))\subseteq
J^{\pm}(\f(\zeta))$ for every set $\zeta\subset V$.
\label{SET}
\end{prop}
\P It is enough to prove it for a single point $p\in V$ and then
getting the result for every $\zeta$ by considering it as
the union of its points. For the first relation, let $y$ be in
$\f(I^{+}(p))$ arbitrary and take $x\in I^{+}(p)$ such that
$\f(x)=y$. Since $p<\!\!<x$ we can
choose a future-directed timelike curve $\g$ from $p$
to $x$. From proposition \ref{curves}, $\f(\g)$ is then a
future-directed timelike curve joining $\f(p)$ and $y$, so that
$y\in I^{+}(\f(p))$. The second assertion is proved in a similar way
using again proposition \ref{curves}.
The proof for the past sets is analogous.\N

This implies that causal relations are ``chronological maps'' in the 
sense of \cite{HARRIS1}. 

\begin{prop}
If $V\prec_{\f} W$ and $\zeta\subset W$ is acausal (achronal) then
$\f^{-1}(\zeta)$ is acausal (achronal).
\label{acausal}
\end{prop}
\P If there were $p,q\in \f^{-1}(\zeta)$ such that $p<q$ ($p\neq q$)
then proposition \ref{SET} would imply $\f(p)<\f(q)$, ($\f(p)\neq
\f(q)$)
with $\f(p),\f(q)\in \zeta$, against the assumption.
And similarly for the achronal case.\N

The impossibility of the existence of causal relations between
given Lorentzian manifolds can be proven sometimes by using results
relating
causality sets or causal curves. The following proposition is an
example.
Let us recall that, for any inextendible causal curve $\g$,
the boundaries $\partial I^{\pm}(\g)$ of its chronological future and
past are usually called its future and past event horizons, sometimes
also called creation and particle horizons, respectively 
\cite{FF,P2,Wald,COND}. Of course these sets
can be empty (then $\g$ has no horizon).
\begin{prop}
Suppose that every inextendible future-directed causal curve in $W$
has a non-empty $\partial I^{-}(\g)$ ($\partial I^{+}(\g)$).
Then any $V$ such that $V\prec W$ cannot have
inextendible causal curves without past (future) event horizons.
\label{HOR}
\end{prop}
\P If there were a future-directed curve $\g$ in $V$ with
$\partial I^{-}(\g)=\emptyset$, $I^{-}(\g)$ would be the whole of
$V$. But according to proposition \ref{SET} $\f(I^{-}(\g))\subseteq
I^{-}(\f(\g))$ from what we would conclude that $I^{-}(\f(\g))=W$ in
contradiction.\N

\subsubsection*{Example 6}\label{ex:x}
\begin{figure}[h]
\epsfxsize=1cm
\hspace{8cm}\epsfbox{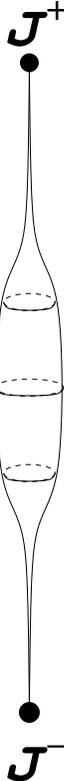}
\caption{Conformal diagram of Einstein spacetime. In this picture the
spacetime is represented by the surface of the figure and each
horizontal circle corresponds to an $(n-1)$-sphere in the Einstein
spacetime. Compare with \cite{Tipler}.}
\label{EINSTEIN}
\end{figure}

Let us recall Example 2 in section \ref{sec:CR}, where we proved that
$\dS\prec\E$, but we did not know if $\E\prec\dS$. Now, by using
proposition \ref{HOR} we have that $\E\not\prec \dS$ because every
causal
curve in the de Sitter spacetime possesses a non-empty event horizon,
see e.g
\cite{FF}, but none of them has one in the Einstein universe.
Thus, $\E\not\sim\dS$.
This is again clear by taking a look at the corresponding conformal
diagrams,
shown in Figures \ref{QUART} and \ref{EINSTEIN}.
Notice that both $\dS$ and $\E$ are locally conformally flat, and therefore
they are
metrically conformally related to each other. However, this conformal
property
is not given by a {\em global diffeomorphism}.

Other impossibilities for causal relations arise from the results
for Cauchy developments.
\begin{prop}
If $V\prec_{\f} W$ then
$D^{\pm}(\f(\zeta))\subseteq \f(D^{\pm}(\zeta))
\,\, \forall \zeta\subseteq V$.
\label{CAUCH}
\end{prop}
\P It is enough to prove the future case. Let $y\in D^{+}(\f(\zeta))$
arbitrary and consider any causal past directed curve
$\g^{-}_{\f^{-1}(y)}\subset V$ containing $\f^{-1}(y)$. Since
$\g^{-}_{\f^{-1}(y)}$ is mapped by $\f$ to a causal curve
passing through $y$, ergo meeting $\f(\zeta)$, we have that
$\g^{-}_{\f^{-1}(y)}$ must meet $\zeta$. As $\g^{-}_{\f^{-1}(y)}$ is
arbitrary we conclude that $y\in\f(D^{+}(\zeta))$. \N
\begin{coro}
If $V\prec_{\f} W$ and $\S\subset W$ is a Cauchy hypersurface then
$\f^{-1}(\S)\subset V$ is a Cauchy hypersurface too.
\label{HYP}
\end{coro}
\P Recall that a Cauchy hypersurface $\S\subset W$ is a closed
acausal set
without edge such that $D(\S)=W$ \cite{BEE,FF,Wald,COND}.
Proposition \ref{CAUCH} implies then
$W=D(\S)\subseteq\f(D(\f^{-1}(\S)))$.
Since $\f$ is a diffeomorphism we get that $D(\f^{-1}(\S))=V$ and that
$\f^{-1}(\S)$ has no edge, so that it only remains to prove its
acausality.
But this is a consequence of proposition \ref{acausal}.\N

\noindent
Let us also recall that a spacetime is {\em globally hyperbolic} if
and
only if it contains a Cauchy hypersurface \cite{BEE,FF,Wald,COND}, see
also
definition \ref{SER} below. Thus we also have
\begin{coro}
If $W$ is globally hyperbolic and $V\prec W$, then $V$ must be
globally
hyperbolic. Thus, if $W$ is globally hyperbolic but $V$ is not,
then $V\nprec W$.\N
\label{HYP2}
\end{coro}
Let us remark that not all diffeomorphic globally hyperbolic
spacetimes are isocausal, as seen for instance in Example 6: $\E\not\sim\dS$.
The last corollaries are very powerful to discard the causal
relationship
between many Lorentzian manifolds. Some outstanding cases are
presented
in the following examples.

\subsubsection*{Example 7}
\begin{figure}[h]
\epsfxsize=2cm
\hspace{8cm}\epsfbox{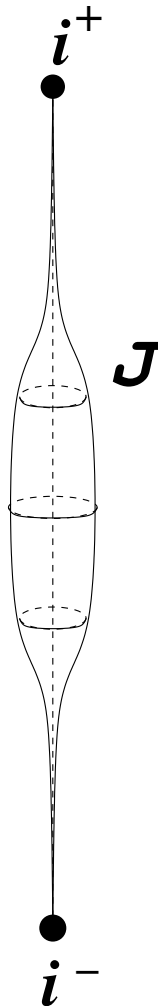}
\caption{Penrose-like diagram for anti-de Sitter spacetime. In this
case we have
preferred to draw a 3-dimensional diagram to get a clearer picture of
the causal
infinity. Every $t=$const.\ slice has been reduced to an open
horizontal disk, so that
every point in the diagram represents a $(n-3)$-sphere except for the
middle line
which is the origin of coordinates. Compare this diagram with that in
\cite{FF}, see also \cite{Tipler}. The boundary of the picture
represents the conformal infinity $\cal{J}$ of $\AdS$. It is
remarkable that this boundary has precisely the shape of the
Einstein universe, see figure \ref{EINSTEIN}. Thus, one is tempted
to say that the causal boundary of $n$-dimensional anti-de Sitter
spacetime is the $(n-1)$-dimensional Einstein universe. We will try
to make the concept of causal boundary precise in section \ref{subsect:CB}.
In any case, notice the timelike
character of $\cal{J}$ and the non-global hyperbolicity of the $\AdS$
spacetime.}
\label{antidS}
\end{figure}

Let us consider anti-de Sitter spacetime $\AdS$: $\r^n$ with
a line-element (in spherical  coordinates $\{t,r,\bar\z_k\}$)
that takes the form
\be
\hspace{-2cm}(\AdS,\T)\, :\, d\tilde{s}^2=\cosh^2 rdt^2-dr^2-
\sinh^2 r\, d\bar{\O}^2_{n-2},\ -\infty<t<\infty,\ 0<r<\infty .
\ee
We compare $\AdS$ with flat spacetime $\L$. By using the standard
spherical coordinates of (\ref{flatsph}) for $\L$ it is very easy to
prove
that the diffeomorphism $\f :\L\rightarrow \AdS$ which identifies
coordinates
in a natural way satisfies  $\f^{*}\T\in\DP^+_2(\L)$, so that
corollary
\ref{pull-back2} implies that $\f$ is a causal relation. Nevertheless,
according to corollary \ref{HYP2}, and since $\L$ is globally
hyperbolic but
$\AdS$ is not (see e.g. \cite{FF} and figure \ref{antidS}, where the
Penrose
diagram of $\AdS$ is shown), we also have that $\AdS\nprec\L$. Hence,
$\AdS\not\sim\L$. Observe that $\AdS$ is locally conformally flat with the
usual
definition, and therefore locally conformally related to $\L$
everywhere.
However, this conformal relation cannot be global, as we have just
proved in a simple way. Therefore, locally conformally flat spacetimes
can have very different causal properties from flat spacetime, and
this can be made precise using the concept of causal relationship.

\subsubsection*{Example 8}
Let us take the particular case of the pp-waves (\ref{ppw}) which are
{\em pure electromagnetic plane waves}: they are locally
conformally flat solutions of the Einstein-Maxwell equations;
for simplicity we take here $n=4$ \cite{KRA}.
These special plane waves are  given by $\ppW (H)$ with
$H=\Phi^2(u)(x^2+y^2)$,
that is
$$
\PW(\Phi): d\bar{s}^2=2d\bar{u}d\bar{v}-
d\bar{x}^2-d\bar{y}^2+\Phi^2(\bar{u})(\bar{x}^2+\bar{y}^2)d\bar{u}^2,\
-\infty<\bar{u},\bar{v},\bar{x},\bar{y}<\infty\, .
$$
As the manifold is $\r^4$ we can try to causally compare these plane
waves
with flat 4-dimensional spacetime $\L$. Defining the usual advanced
and retarded
null coordinates the line-element for $\L$ can be written as
$$
\L :ds^2=2dudv-dx^2-dy^2\, .
$$
Using  the diffeomorphism given by
$\bar{u}=u,\ \bar{v}=v,\ \bar{x}=x,\ \bar{y}=y$,
a calculation analogous to that of the Example 5 proves that $\f$ is
a causal
relation with $\vec k$ as canonical null direction. Nevertheless, for
all
$\Phi \neq 0$ the plane waves $\PW (\Phi)$ are known to be
non-globally hyperbolic \cite{WAVE} and hence the causal relation in
the
opposite way is not possible. Thus, for all $\Phi\neq 0$,
$\L\not\sim\PW(\Phi)$.
Observe that again all the $\PW(\Phi)$ spacetimes are locally
conformally flat,
but this does not mean that they are isocausal to $\L$,
which is a global property.

\subsection{Causally ordered sequences of Lorentzian manifolds.
Causal structures}
Globally hyperbolic spacetimes are the best-behaved Lorentzian
manifolds from
the causal point of view and, as we have seen in corollary \ref{HYP2},
if $W$ has this property and is causally related to $V$, then $V$ must
also have it. We can then ask ourselves whether other milder causality
conditions behave in a similar way under causal relations. To that
end,
let us briefly recall here the standard hierarchy of causality
conditions
\cite{COND,FF}.
\begin{defi}
A Lorentzian manifold $V$ is said to be:
\begin{itemize}
\item {\bf not totally vicious} if $I^+(p)\cap I^-(p)\neq V\ \forall
p\in V$.
\item {\bf chronological} if $p\not\in I^{+}(p)\ \forall p\in V$.
\item {\bf causal} if $J^{+}(p)\cap J^{-}(p)=\{p\}\ \forall p\in V$.
\item {\bf future distinguishing} if $I^{+}(p)\neq I^{+}(q)$ $\forall
p\neq q$,
and analogously for the past. This is equivalent to demanding that
every neighbourhood of $p$ contains another neighbourhood $\U_p$ of
$p$
such that every causal future directed curve starting at $p$ intersects $\U_p$ in a connected set.
\item {\bf strongly causal} if $\forall p\in V$ and for every neighbourhood $\W_p$ of $p$ there exists another neighbourhood
$\U_p\subset\W_p$ containing $p$ such that for every causal curve $\g$ the intersection $\g\cap\U_p$ is either empty or a connected set.
\item {\bf causally stable} if there exists a function whose gradient
is
timelike everywhere (called a time function).
\item {\bf globally hyperbolic} if it is strongly causal
and $J^{+}(p)\cap J^{-}(q)$ is compact for all $p, q\in V$.
\end{itemize}
\label{SER}
\end{defi}
These conditions are given with increasing degree of restriction so
that any of
them implies all the previous. The next result proves that these
constraints are kept  by causal relations.
\begin{theo}
Let $V\prec W$. Then, if $W$ satisfies any of the causality
conditions of
definition \ref{SER}, so does $V$.
\label{INCR}
\end{theo}
\P Let $V\prec_{\f} W$. We prove each case separately.
If $V$ were totally vicious there would be
a $p\in V$ such that $I^+(p)=I^-(p)=V$, so that from proposition
\ref{SET}
$W=\f(I^{\pm}(p))\subseteq I^{\pm}(\f(p))$ and thus
$I^+(\f(p))\cap I^-(\f(p))=W$ proving that $W$ would be totally
vicious, against
the hypothesis.

Suppose $V$ were not a chronological spacetime.
Then, from proposition \ref{curves} $\f$ would map every closed
timelike
curve of $V$ onto a closed timelike curve of $W$, so that $W$ could
not
be chronological. The proof for a causal spacetime $W$ is similar.

Suppose now that $V$ were not future distinguishing. Then, there
would be
a point $p\in V$ and a neighbourhood $U_p$ of $p$ such that every
open set $\U_p$
with $p\in\U_p\subset U_p$ would cut at least a causal curve
$\g$ starting at $p$ in a disconnected set $\g\cap\U_p$. But then, using
proposition \ref{curves} again, every open subset of $\f(U_p)$ would
also cut
the causal curve $\f(\g)$, which starts at $\f(p)$, in a
disconnected set,
hence $W$ would not be future distinguishing. The past case is
identical. The
proof for the strongly causal spacetimes is also similar.

Now, let $W$ be causally stable, and let $\t$ be the function such
that
$d\t$ is an everywhere timelike and future directed 1-form. By
proposition
\ref{CAUS} point $(iii)$, $\f^{*}(d\t)=d(\f^{*}\t)$ is also a
future-directed
timelike 1-form in $V$, and hence $\f^{*}\t$ is the required time
function
for $V$. Finally, the globally hyperbolic case is corollary
\ref{HYP2}.\N

We proved in section \ref{sec:CR} that the relation $\prec$ is a
preorder,
hence $(\cal{L},\prec)$, where $\cal{L}$ denotes the class of all
Lorentzian manifolds, is a preordered set. Of course,
$\prec$ only
pre-orders the Lorentzian manifolds which are pairwise diffeomorphic,
so that
in fact each of the subsets $\mbox{Lor}(M)$ are in fact separately
preordered by
$\prec$, where $\mbox{Lor}(M)$ denotes the set of Lorentzian manifolds
with base
manifold $M$. As usual, the equivalence
relation constructed from $\prec$, which is the ``$\sim$''
providing the definition
\ref{EQUIV} of isocausal spacetimes, gives rise to a partial order
in the quotient sets $\mbox{Lor}(M)/\!\!\sim$ by means of the new
binary relation
$\mbox{coset}(V)\preceq \mbox{coset}(W) \Leftrightarrow V\prec W$.
Here
coset$(V)=\{U \ :\, V\sim U\}$ denotes the equivalence class  of
spacetimes
isocausal to $V$.

All this means that $\cal{L}$, and in fact each of the $\mbox{Lor}(M)$,
can be decomposed in disjoint and
partially ordered classes of isocausal Lorentzian manifolds.
Of course, still we may find classes coset$(V_1)$ and coset$(V_2)$
belonging
to $\mbox{Lor}(M)/\!\!\sim$
which are not related by $\preceq$ at all. Nevertheless, it is in
principle
possible to construct causally ordered sequences of spacetimes in
which every
pair of elements of the sequence are comparable with respect to the
binary
relation $\preceq$. These sequences look like
\be
\mbox{coset}(V)\preceq\dots\preceq\mbox{coset}(W)\preceq\dots\preceq
\mbox{coset}(U)\preceq\dots\preceq\mbox{coset}(Z)
\label{sequence}
\ee
where, from theorem \ref{INCR}, if a member of the sequence
satisfies one
of the causality conditions of definition \ref{SER}, then all the
previous
members (those to the left) comply also with the same condition;
and reciprocally, if one of them violates one of those conditions,
then
all the members to the right violate it too.
Since the causality conditions of definition \ref{SER} are given with
increasing order of restriction, we deduce that spacetimes which have
stronger
causality properties appear towards the left of the sequence,
whereas spacetimes with weaker causality conditions appear towards
the right of
(\ref{sequence}). All this is quite natural because the Lorentzian
cones
open up under a causal relation. It also provides an abstract measure
of
``increasing causality'': the ``smaller'' the spacetime in a sequence,
the better causal behaviour it has.

The longest sequences of type
(\ref{sequence}) are those starting with a simple globally hyperbolic
spacetime
(say a flat space such as $\L$, or equivalently any of the members in coset$(\L)$,
such as $\fr{1}{4}\dS$ or $\RW_0\{\fr{3-n}{n-1}\}$), passing through a $W$ which
is causally
stable (say anti de Sitter $\AdS$, as $\L\prec \AdS$), and so on
until they
end with a causally rather badly behaved Lorentzian manifold.
Of course, there can be various steps in a sequence with a given
property of
definition \ref{SER} (for instance, not all diffeomorphic
globally hyperbolic spacetimes are isocausal, e.g., $\E \not\prec \dS$):
thus the binary relation $\preceq$ is finer than the classification of
definition \ref{SER}.  Whether or not the last step in these longest
sequences is
always a totally vicious
spacetime\addtocounter{footnote}{-5}\footnote{Totally
vicious spacetimes do exist and may be quite simple: one example is
the famous G\"odel spacetime \cite{Godel,FF}. Another will be 
presented in Example 11.}, which would provide a
maximal element to the partial order $\preceq$, is an interesting open
question. Another question is if there is a minimal element for each
sequence, providing the ``best'' causally behaved spacetime for a
given
manifold.

All the Lorentzian manifolds involved in a given sequence
of type
(\ref{sequence}) are diffeomorphic to each other, as they belong to
$\mbox{Lor}(M)$
and therefore all of them are diffeomorphic to $M$. Consequently,
perhaps
a more interesting way to look at the previous results is to
consider all the classes of equivalence of spacetimes in a sequence
as different
causal structures on the {\em same} manifold $M$. More precisely
\begin{defi}
Let $M$ be a differentiable manifold. A {\bf causal structure} on $M$
is an equivalence class  with respect to $\sim$ of Lorentzian
manifolds based at $M$.
\label{STRUC}
\end{defi}
Of course, not all
manifolds possess a causal structure, for as is well-known not every
differentiable manifold possesses a global Lorentzian metric (take
for instance
$S^n$). On the other hand, there are manifolds with many inequivalent
causal
structures such as for example $\r^n$: just consider coset$(\L)$,
or coset$(\AdS)$, or the equivalence class  of G\"{o}del spacetime.
Therefore, for any given differentiable manifold admitting causal
structures,
these can be partially ordered according to $\preceq$
and we can construct sequences of type
(\ref{sequence}). Interesting open questions are the cardinality of
the possible inequivalent causal structures admitted by a given manifold,
and the possible existence of minimal and maximal elements.

According to the definition \ref{STRUC}, two Lorentzian metrics
$\G_1$ and $\G_2$ on $M$ are said to be equivalent from the causal point 
of view if the Lorentzian manifolds
$(M,\G_1)$ and $(M,\G_2)$ are isocausal.
In other words, effectively a causal structure on $M$ is simply
coset$(V)\subset \mbox{Lor}(M)$ with {\em any} of its Lorentzian
metrics. Note that specific metric
properties (distances, proper times, volumes, etcetera) are
completely irrelevant
here. An important remark is that our definition of causal
structure is more general than the traditional ``conformal'' one.
If one adopts definition \ref{STRUC}, then the global causal structure
of a given Lorentzian manifold is {\em not} given up to
a conformal factor of the metric. Rather, it only determines
coset$(V)$, i.e., {\em the metric up to causal
mappings.} Whether or not this generalization is adequate depends on 
the type of properties one wishes to keep. For instance, it is 
intuitively clear that the causal structure of a weak static gravitational 
field far from the sources should be similar to that of flat 
spacetime $\L$. However, no realistic gravitational field will be 
conformal to $\L$, not even far from the sources. Thus, the conformal 
structure does not capture the intuitive concept that these two 
situations share somehow the same causality properties. As we are 
going to prove in the next examples, the generalization given by 
definition \ref{STRUC} provides a rigorous framework, and a 
justification, for that intuitive claim.

\subsubsection*{Example 9}\label{schw}
Consider the outer region of $n$-dimensional Schwarzschild spacetime
$\Sc$
in typical spherical coordinates, whose line-element for positive
mass $M$ reads
\be
d\tilde{s}^2=\left(1-\fr{2M}{r^{n-3}}\right)dt^2-
\left(1-\fr{2M}{r^{n-3}}\right)^{-1}dr^2 -r^2d\bar{\O}^2_{n-2}
\ee
and take the Lorentzian manifolds $\Sc_c$ defined
by $-\infty<t<\infty$ and the condition
$r>c\geq (2M)^{\fr{1}{n-3}}\equiv c_M$.
The second spacetime is $\L$ in spherical
coordinates as in (\ref{flatsph}) of Example 3, but in order to make it
diffeomorphic with
$\Sc_c$ we need to take only a subregion $\L_a$
defined by the condition $R>a$ for a fixed non-negative constant $a$.

We want to study the causal relationship between $(\Sc_c,\T)$ and
$(\L_a,\G)$.
To than end, and to avoid unnecessary writing, we will omit the
angular coordinates in what follows, as they are simply identified for
all diffeomorphisms under consideration. Define first
$\f_b:\L_a\rightarrow \Sc_c$
by $t=bT,\ r=R-a+c$ where $b$ is a positive constant. A simple
computation
provides the eigenvalues of $\f_b^*\T$ with respect to $\G$, given by
$$
b^{2}\left(1-\fr{2M}{(R-a+c)^{n-3}}\right),\,
\left(1-\fr{2M}{(R-a+c)^{n-3}}\right)^{-1},\,\mbox{and}\,\,
\left(\fr{R-a+c}{R}\right)^{2}.
$$
Thus if $\f_b^{*}\T$ is to be in $\DP^{+}_2(\L_a)$, according to
criterion
\ref{ORTHONORMAL} the following inequalities must hold:
$$
b^{2}\left(1-\fr{2M}{(R-a+c)^{n-3}}\right)\geq\left\{
\left(1-\fr{2M}{(R-a+c)^{n-3}}\right)^{-1},\
\left(\fr{R-a+c}{R}\right)^{2}\right\}.
$$
These can be satisfied for every $c>c_M$ by arranging
$b$ appropriately. Hence, according to corollary \ref{pull-back2} we
deduce $\L_a\prec \Sc_c$ for all $c>c_M$.

Reciprocally, let $\psi :\Sc_c\rightarrow \L_a$ be defined simply by
means of
$T=t,R=r$. Then the eigenvalues of $\psi^{*}\G$ with respect to $\T$
are
given by $\left(1-\fr{2M}{r^{n-3}}\right)^{-1}$, $1-\fr{2M}{r^{n-3}}$
and $1$.
Criterion \ref{ORTHONORMAL} implies that
$\psi^{*}\G\in\DP^{+}_2(\Sc_c)$
for every $c\geq c_M$ as long as
$a\geq c_M$, and corollary \ref{pull-back2} leads to
$\Sc_c\prec \L_c$ for all $c\geq c_M$. The conclusion is that $\Sc_c\sim\L_c$
if $c>c_M$, as was to be expected. This example can be repeated for a global
spacetime $\hat\Sc_{c}$ ---formed by Schwarzschild exterior matched to some
adequate interior at $c>c_{M}$---
and the whole of Minkowski spacetime $\L$. The two manifolds are then 
diffeomorphic. The conclusion again is that $\hat\Sc_{c}\sim \L$ if $c>c_M$.

Notice that $\Sc_{c}$ is not locally conformally flat, and therefore
is not conformal to $\L_{c}$.
This means that the conformal structure does not allow to say that 
$\Sc_{c}$ and $\L_{c}$ have a similar causality, while the 
concept of isocausality certainly does, at least up to a point, because 
$\Sc_{c}\in$ coset$(\L_{c})$. Since $\Sc_{c}\sim\L_{c}$,  
some causal features are shared by these two spacetimes, but of course
not {\em all} thinkable causal properties. For instance, in $\Sc$ there are
circular null geodesics at $r=[(n-1)/2]^{\fr{1}{n-3}}c_{M}$, but there are
clearly none in $\L$. A more drastic example is given by the following 
property \cite{P3}: for all endless causal curves $\g_{1}$ 
and $\g_{2}$ in $\Sc$, $I^+(\g_{1})\cap \g_{2}\neq \emptyset$, and 
similarly for the past. This could be termed as a causal property, 
but it is not shared by $\L$, as there are some simple examples in 
Minkowski spacetime of endless timelike curves which are completely
causally disconnected, see \cite{P3}. In a way, this is a consequence 
of the existence of a gravitational field in $\Sc$, maybe weak, but 
non-vanishing nonetheless. Such kind of properties could only be kept by 
the fully faithful conformal structure, but then one would lose the 
possibility of giving a meaning to the intuitive concept of having 
close-to-Minkowskian causality in weak fields far from the sources.

\begin{figure}
\epsfxsize=6cm
\hspace{6cm}\epsfbox{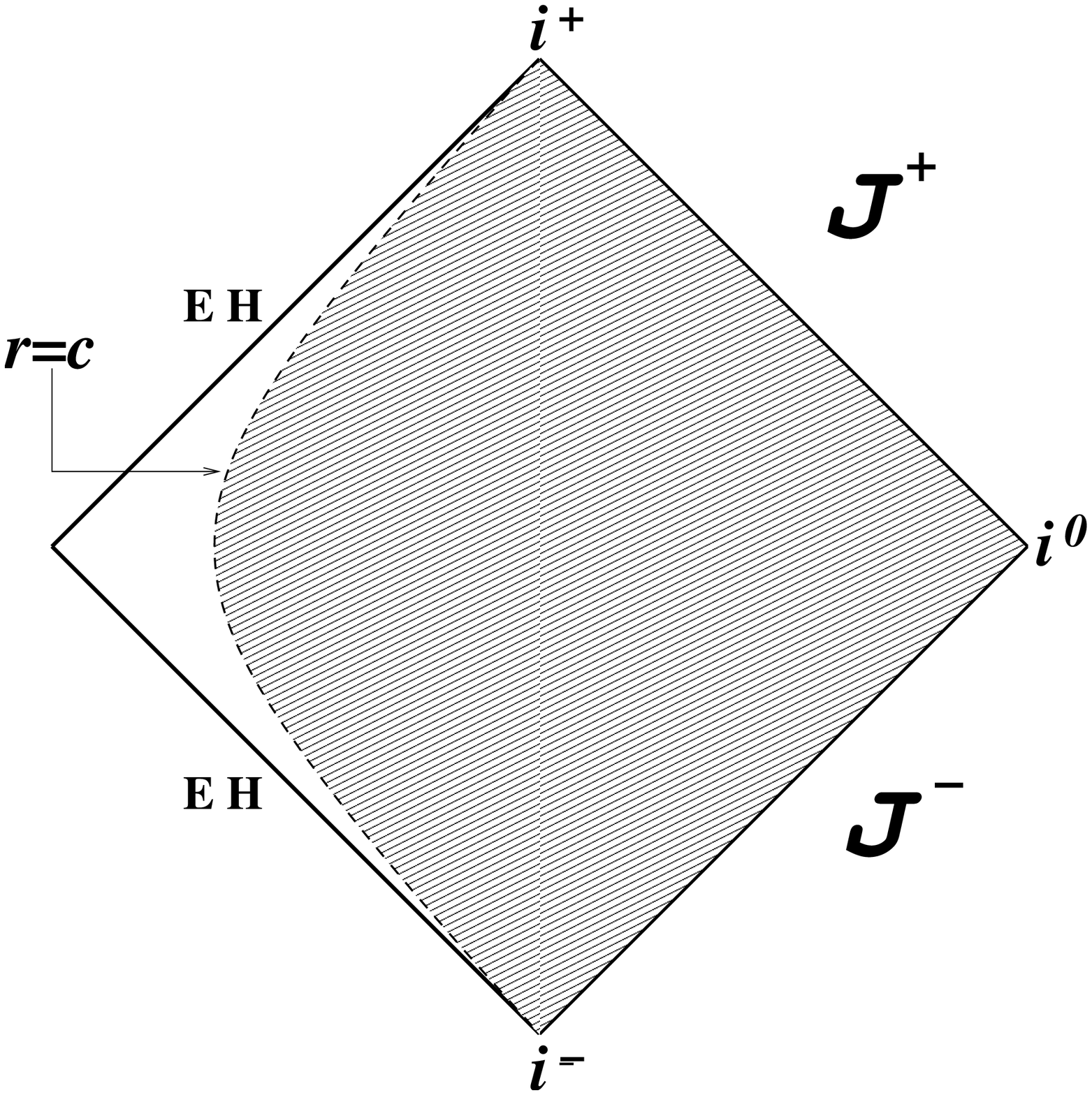}
\caption{Penrose diagram of the exterior region of Schwarzschild
spacetime.
Only the
part to the right of $r=c$ (shaded zone) is the spacetime $\Sc_c$
which is
isocausal to $\L_c$. In this case, as is clear from the figure,
$\Sc_c$ is not
globally hyperbolic. On the other hand, the limit case with $c=c_M$,
in which the
boundary of the spacetime is the event horizon EH, is globally
hyperbolic. This
spacetime is {\em not} isocausal to any part $\L_a$ of flat
spacetime.}
\label{SCHWARZ}
\end{figure}

 With isocausality we have kept, for instance, the causal stability of 
both $\Sc_{c}$ and $\L_{c}$, or the global hyperbolicity of 
$\hat\Sc_{c}$ and $\L$, in a precise mutual way.
This can be highlighted by noting the following
remark, which also has 
physical implications: we have not proved $\Sc_c\sim\L_c$ for the
extreme value $c=c_M$, since $\f_b$ failed to be causal relations in that
case. In fact, we can disprove $\L_a\prec \Sc_{c_M}$ $\forall a\geq 0$
by making use of corollary \ref{HYP2}, because $\Sc_{c_M}$ is globally
hyperbolic but $\L_{a\geq 0}$ is not (recall also that the manifolds
$\Sc_c$ and $\L$ are not diffeomorphic). We conclude then
that $\L_a\not\prec \Sc_{c_M}$. This is a very interesting result,
being a clear manifestation of the null character of the event horizon
$r=c_M$ in the extensions through it of Schwarzschild's spacetime.
It is remarkable that we have not made direct use of any extension or
extendibility of $\Sc_{c_M}$ to achieve this result (although we have
clearly used its global hyperbolicity).
Once again, a clear picture of what is happening can be obtained by
taking a look at the Penrose diagrams, corresponding to the part to the
right of $r=a$ in figure \ref{Minkowski} for $\L_a$ and to the one
presented in figure \ref{SCHWARZ}. Another example of this type is
provided next.

\subsubsection*{Example 10}\label{Sch-RN}
In this example we will prove that the outer regions of
Schwarzschild ($\Sc$) and Reissner-Nordstr\"om ($\RN$) black holes in
$n$ dimensions are isocausal. In order to avoid complications arising
from the parameters which appear in the line element of these
spacetimes, we will use spherical dimensionless coordinates
in which the line elements take the form:
\begin{eqnarray*}
(\Sc,\G) :
ds^2=\bar{\alpha}^2
\left[\left(1-\fr{1}{r^{n-3}}\right)dt^2-
\left(1-\fr{1}{r^{n-3}}\right)^{-1}dr^2-r^2d\O^2_{n-2}\right],\\
\hspace{-25mm}(\RN,\T):
d\tilde{s}^2=\alpha^2
\left[\left(\fr{1}{r_0^{n-3}}-\fr{1}{R^{n-3}}\right)
\left(\fr{1}{r_1^{n-3}}-\fr{1}{R^{n-3}}\right)dT^2-\right.\\
-\left.\left(\fr{1}{r_0^{n-3}}-\fr{1}{R^{n-3}}\right)^{-1}
\left(\fr{1}{r_1^{n-3}}-\fr{1}{R^{n-3}}\right)^{-1}dR^2-R^2d\bar{\O}^2_{n-2}
\right],
\end{eqnarray*}
where $\alpha=Q$ is the charge of $\RN$ and $\bar{\alpha}=2M$ is the
mass of $\Sc$ (we have arranged the original metrics of both black
holes in such a way that $M$ and $Q$ have dimensions of length). The
parameters $r_0$ and $r_1$ correspond respectively to the usual Cauchy
and event horizons $r_-$ and $r_+$ of the $\RN$ black hole by means of the
relations $r_-=Qr_0$ and $r_+=Qr_1$. We are only interested here in
the outer regions of both spacetimes, that is to say, $r>1$ for $\Sc$
and $R>r_1$ for $\RN$, which are globally hyperbolic.
These regions are covered by the previous
coordinate systems with the time coordinates $t$ and $T$ running over
the whole real line. It is then very easy to write down
diffeomorphisms which set up the mutual causal relation. Omitting the
angular variables as before, we can choose $\f_b :\Sc \rightarrow \RN$
defined by $T=bt,\  R=r_1r$ and $\psi_a :\RN\rightarrow \Sc$
by  $t=aT,\ r=R-r_1 +1$. A calculation similar to those performed in
previous
examples, and use of either of the criteria \ref{cri:1} or
\ref{ORTHONORMAL},
allows us to find the conditions for the tensors $\f^{*}_b\T$ and
$\psi^{*}_a\G$ to be causal:
\begin{eqnarray*}
\left(\fr{1}{r_0^{n-3}}-\fr{1}{r^{n-3}r_1^{n-3}}\right)
\geq\left\{\fr{r_1^{n-2}}{b},\
\fr{r^{n-1}_1}{b^2}\right\} \, \Longleftrightarrow\,
\f^{*}_b\T\in\DP^+_2(\Sc),\\
\hspace{-20mm}a\left(1-\fr{1}{(R-r_1+1)^{n-3}}\right)
\geq\left\{\left(\fr{1}{R^{n-3}}-\fr{1}{r_0^{n-3}}\right)
\left(\fr{1}{R^{n-3}}-\fr{1}{r_1^{n-3}}\right)\right.,\\
\hspace{-20mm}\left.\fr{1}{a}\left(1+\fr{1-r_1}{R}\right)^2
\left(\fr{1}{R^{n-3}}-\fr{1}{r_0^{n-3}}\right)\left(\fr{1}{R^{n-3}}-\fr{1}{r_1^{n-3}}\right)\right\}
\, \Longleftrightarrow\, \psi^{*}_a\G\in\DP^+_2(\RN).
\end{eqnarray*}
It is not difficult to see that these conditions are complied for
suitable values of the parameters $a$ and $b$. Therefore, from
corollary
\ref{pull-back2} we obtain $\Sc\sim\RN$. This was to be expected
since the Penrose diagrams of the considered regions of these
two spacetimes have the same shape.

\subsection{Future and past objects}
Let us now pass to the question of how future and past objects
transform under a
causal relation. It is enough to concentrate on the future case but
clearly
all the statements have a counterpart for the past which we will
sometimes
make explicit. The results for future tensors and future-directed 
curves were given in propositions \ref{CONSERVATION} and 
\ref{curves}, respectively. For future sets we have
\begin{prop}
If $V\prec_{\f} W$ then $\f^{-1}(\B^+)$ is a future set for every
future set
$\B^+\subseteq W$.\label{KEY}
\end{prop}
\P Suppose that $V\prec_{\f} W$, that $\B^+\subseteq W$ is a future
set,
and take  $\f^{-1}(\B^+)\subseteq V$. Proposition \ref{SET} implies
$\f(I^{+}(\f^{-1}(\B^+)))\subseteq
I^{+}(\f(\f^{-1}(\B^+)))=I^{+}(\B^+)\subseteq\B^+$ proving that
$I^{+}(\f^{-1}(\B^+))\subseteq \f^{-1}(\B^+)$.\N
\begin{prop}
If $\B\subset W$ is an achronal boundary and $V\prec_{\f} W$
then $\f^{-1}(\B)$ is also an achronal boundary in $V$.
\label{ACHR-BOUND}
\end{prop}
\P If $\B\subset W$ is an achronal boundary then by definition
there is a future set $\B^{+}$ such that $\B=\d\B^+$. Since $\f$ is a
diffeomorphism we have $\f^{-1}(\B)=\f^{-1}(\d\B^+)=\d(\f^{-1}(\B^+))$
\cite{DUGUN}. This proves, on account of proposition \ref{KEY}, that
$\f^{-1}(\B)$ is the achronal boundary of the future set
$\f^{-1}(\B^+)$.\N

It can be shown that every achronal boundary is an embedded
$(n-1)$-dimensional
$C^{1^-}$ hypersurface without boundary \cite{BEE,FF,COND,Wald}.
Proposition \ref{ACHR-BOUND} tells us that the achronality of this
particular
kind of hypersurfaces is preserved under $\f^{-1}$ for a causal $\f$, and
proposition
\ref{CAUCH} proved that the property of being a Cauchy hypersurface
is also
preserved by $\f^{-1}$.

Propositions \ref{KEY}, \ref{ACHR-BOUND}, \ref{CONSERVATION} and
\ref{curves} can be 
combined to prove the existence of bijections between the future 
objects of isocausal spacetimes. We collect this in the following 
corollary.
Let us denote by ${\cal F}_V$ and ${\cal F}_W$ the classes of future
sets of $V$ and $W$, respectively.
\begin{coro}
Let $V\sim W$. Then ${\cal F}_V$ and ${\cal F}_W$ have the same 
cardinality, and similarly for the past sets, the causal curves and
the proper achronal boundaries of $V$ and $W$.
\label{CONSERV}
\end {coro}
\P If $V\sim W$ then $V\prec_{\f}W$ and $W\prec_{\psi}V$ for some
diffeomorphisms $\f$ and $\psi$. Now, due to proposition \ref{KEY},
$\f^{-1}({\cal F}_W)\subseteq{\cal F}_V$ and
$\psi^{-1}({\cal F}_V)\subseteq{\cal F}_W$. Since both $\f$ and $\psi$
are bijective maps we conclude that ${\cal F}_V$ is in one-to-one
correspondence with a subset of ${\cal F}_W$
and vice versa which, according to the equivalence theorem
of Bernstein \cite{BERN}, implies that ${\cal F}_V$ is in one-to-one
correspondence with ${\cal F}_W$. The rest of the cases are proved 
analogously. \N

{\it Remark} 
The cardinality of the set of causal curves in {\em any} Lorentzian 
manifold is that of the continuum, so that this corollary is trivial 
for future-directed causal curves, and also for future tensor 
{\em fields}, that is to say for the sections of $\DP^{\pm}(V)$. However,
we are regarding here $\DP^{\pm}_r(V)$ as a subset of
the bundle $T(V)$.  The matter is not quite so simple regarding future 
and past sets, and achronal boundaries, as the cardinality of, say, ${\cal F}_V$
varies for different $V$. Of course, in any future-distinguishing 
spacetime the cardinality of ${\cal F}_V$ is, {\em at least}, that 
of the continuum. But for non-distinguishing spacetimes this can change 
drastically. For example, if $V$ is a totally vicious spacetime,
then $I^+(x)=I^-(x)=V$ for all $x\in V$, see
e.g. proposition 2.18 in \cite{COND}, hence such a $V$ 
contains just one future (and past) set, namely the manifold $V$
itself, and no proper achronal boundaries. Therefore, according to corollary
\ref{CONSERV} these spacetimes cannot be
isocausal to a non-totally vicious spacetime (this can also
be seen from theorem \ref{INCR}). Other possibilities are shown in 
Example 11 below.

One wonders if the future sets and their properties may serve as basic
objects in order to construct the causal structure of a spacetime 
without using the conformal metric. This would be 
analogous to what happens in topology with open sets, which are enough to
build up all the usual topological concepts such as
continuity, compactness, etcetera, making no use of further structures as those
introduced when a notion of distance is defined.
 From proposition \ref{KEY} and corollary \ref{CONSERV} we know that once we
have defined the future and past sets in a Lorentzian manifold,
we cannot put the future sets and past sets in another isocausal manifold
arbitrarily. This is somehow reminiscent of the Geroch, Kronheimer and Penrose
(henceforth GKP) definition of causal boundary--- see section 
\ref{subsect:CB}--- for distinguishing 
spacetimes, where the whole scheme is based on the so-called IF's 
(irreducible future sets) and their past counterparts, see \cite{GKP,FF}.

\subsubsection*{Example 11}
\begin{figure}[h]
\epsfxsize=16cm
\epsfbox{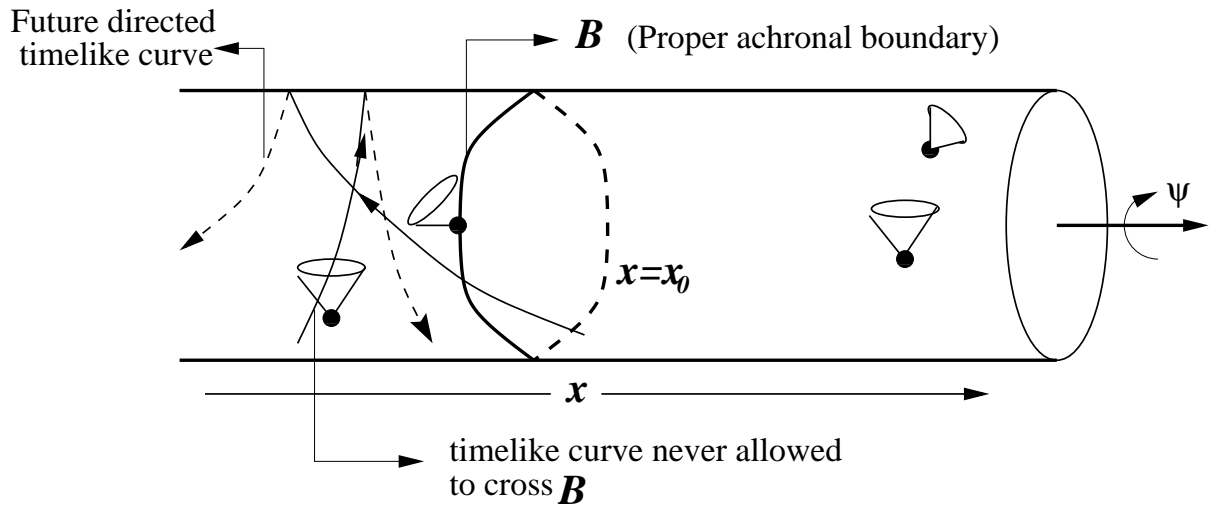}
\caption{This is an schematic picture of the spacetime analyzed in Example 11.
The shown null curve 
$x=x_{0}$ is an achronal boundary and the future of any point to the 
left of $B$ is the region $x< x_{0}$, so that $x< x_{0}$ is a future set.
Analogously, $x>x_{0}$ is a past set.}
\label{FUTURE}
\end{figure}

It seems clear that totally vicious spacetimes are the worst causally behaved
spacetimes since they have just one future (past) set and no proper achronal
boundaries.  The following step in the causality ladder should be
the spacetimes with a finite number of causal sets.  Examples of
such spacetimes are given by the following line-element:
$$
(V,\G):ds^2=-2f^2(x)d\psi dx+g^2(x)d\psi^2-dx^2,\ -\infty<x<\infty,\ 
0<\psi <2\pi.
$$
This is a two-dimensional spacetime with $\r\times S^1$ as base
manifold. We assume that the functions $f(x)$ and $g(x)$ have no common zeros
so as to have det\,$\G\neq 0$. The vector $\fr{\d}{\d \psi}|_{x=x_0}$
is null at each zero $x_0$ of $g(x)$ generating thus a closed null
curve $\g$ diffeomorphic to $S^1$.  In fact, any such $\g$ is a proper
achronal boundary and acts as a one-way membrane for the timelike
future-directed curves moving towards decreasing values of $x$. 
Therefore if we pick up a point $p\in V$ such that $x(p)<x_0$ we get that
for every $q\in I^+(p)$, $x(q)<x_0$ (see figure \ref{FUTURE}). 
Another way of looking at this is that the future null cone at
each point of $\g$ is tilted towards negative values of $x$. This can be
explicitly worked out by considering any vector $(a_1\fr{\d}{\d x}
+a_2\fr{\d}{\d \psi})|_{x=x_0}$ and requiring it to be 
future-directed (so that $a_2>0$) and timelike (which implies then that 
$a_1$ must be negative.)  A similar reasoning replacing future by past 
leads to the corresponding conclusions for past-directed curves
with $\g$ acting now as a one-way membrane in the opposite direction.

It follows that $V$ has as many proper achronal boundaries as the 
number of zeros of $g(x)$, which may be finite or infinite countable,
and the number of future and past sets is that number plus one. If 
$g(x)$ has no zeros, then the spacetime is totally vicious. If it has 
$m$ zeros, then the $m+1$ future sets are nested, in the sense that 
all the future sets whose achronal boundary is $x=x_{j}$ are proper 
subsets of the future sets whose achronal boundary is $x=x_{i}$ 
with $x_{j}< x_{i}$.

According to corollary \ref{CONSERV}, any spacetime with a 
finite number of achronal boundaries (or future sets), as those shown 
in this example,  can only be isocausal to 
spacetimes with exactly the same number of achronal boundaries (or 
future sets).

\subsection{Sufficient conditions for a causal relationship}
We have proved the interesting propositions \ref{curves}, \ref{SET},
\ref{acausal}, \ref{CAUCH}, \ref{KEY}, \ref{ACHR-BOUND}, but apart
from the
first one all the rest provide only necessary conditions for a
diffeomorphism
to be a causal relation. Now we are going to present the appropriate
sufficient conditions by proving partial converses of some of these
results. In order to see that these converses cannot be so simple let
us
start with an illustrative result of how some ``natural'' sufficient
conditions
may fail to work. Consider for instance the condition found in
proposition \ref{SET}.
\begin{lem}
Let $\f:V\rightarrow W$ be a diffeomorphism with the property
$\f(I^{+}(p))\subseteq I^{+}(\f(p))\ \forall p\in V$. Then, for all
timelike future-directed curves $\g\subset V$, any two points
$x,y\in \f(\g)$ satisfy $x<< y$ or $y<< x$.
\label{CHRONOLOGICAL}
\end{lem}
\P Take any future-directed timelike $\g\subset V$ and any two points $p,q\in
\g$, so that
$p<< q$. It is clear that if the assumption for $\f$ holds, then
$\f(p)<\!\!<\f(q)$.\N

Still, the conclusion of this lemma does {\em not} imply that
$\f(\g)$ is a timelike curve, even though all its points are chronologically
related. Explicit examples of the opposite are given by all totally vicious
spacetimes $W$, in which all curves (be them causal or not) satisfy the
property that $x<<y$ for every pair of its points.
And of course there are spacelike curves in $W$.

What we need here to avoid
these counterexamples is to require some causal property for the
spacetime $W$.
\begin{lem}
Let $V$ be a future and past distinguishing Lorentzian manifold.
Then every curve $\g$ satisfying that $p<<q$ or $q<< p$ for all
$p,q\in\g$ is timelike and causally oriented.
\label{dist}
\end{lem}
\P Pick up any $p\in \g$ and let $\NN_p$ be a normal neighbourhood of
$p$. As $V$ is future and past distinguishing there is another neighbourhood
$\U_p\subset\NN_p$ of $p$ such that {\em all} causal curves starting at $p$ cut
$\U_p$ in a connected set. Choose any $z\in\g\cap \U_p$, so that by
assumption $z<<p$ or $p<<z$. In the second possibility there is a timelike
future-directed segment $\l$, with past and future endpoints at $p$ and $z$
respectively, such that $\l\cap \U_p$ must be connected.
This implies that $\l\subset \U_p$ as $\U_p$ is open, hence
$\l$ is a future-directed timelike segment contained in the normal
neighbourhood $\NN_p$. And similarly, but past-directed, in the other
possibility $z<<p$. As $z$ was arbitrary, such a segment can thus be
constructed for all $z\in\g\cap \U_p\subset\NN_p$, which implies 
that $\g$ is timelike nearby $p$. 
Covering $\g$ with sets of the type $\g\cap \U_p$, $p\in\g$,
the result follows.\N

Now we can prove an important partial converse to proposition
\ref{SET}.
\begin{prop}
Let $W$ be future and past distinguishing and $\f:V\rightarrow W$
a diffeomorphism such that $\f(I^{+}(p))\subseteq I^{+}(\f(p))\
\forall p\in V$.
Then $\f$ is a causal relation and, as a consequence, $V$ is also
future
and past distinguishing.
\label{SET2}
\end{prop}
\P Take any future-directed timelike curve $\g\subset V$.
 From lemma \ref{CHRONOLOGICAL} we have that $x<< y$ or $y<< x$
for all $x,y\in \f(\g)$, and then lemma \ref{dist} implies that
$\f(\g)\subset W$ is a future-directed timelike curve. As $\g$ was
arbitrary, proposition \ref{curves} tells us that $\f$ is a causal
relation,
and then theorem \ref{INCR} ensures that $V$ must be
distinguishing.\N

Finally, we can also prove partial converses to propositions
\ref{KEY} and \ref{ACHR-BOUND}, which are key results in our work.
But first we need a simple lemma taken from \cite{PENROSE}.
\begin{lem}
If $\B^+$ is a future set then
$p\in\overline{\B^+}\Longleftrightarrow I^{+}(p)\subseteq\B^+$.
\label{CLOSURE}
\end{lem}
\P It is well-known that, for {\em any} set $\zeta$,
$\overline{I^+(\zeta)}=\{x\in V :\ I^+(x)\subseteq I^+(\zeta)\}$, see
e.g. point (iv) in proposition 2.15 of \cite{COND}. But for a future set
$\overline{\B^+}=\overline{I^+(\B^+)}$, from where the result
follows.\N
\begin{theo}
Let $W$ be future and past distinguishing. Then, a diffeomorphism
$\f : V\rightarrow W$ is a causal relation if and only if
$\f^{-1}(\B^+)$ is a future set for every future set
$\B^+\subseteq W$. And similarly for the past. \label{KEY2}
\end{theo}
\P One implication is proposition \ref{KEY}. For the converse, take
any
$p\in V$ and the future set $I^{+}(\f(p))$. Due to the assumption,
$\f^{-1}(I^{+}(\f(p)))$ is a future set.
Since $\f(p)\in\overline{I^{+}(\f(p))}$ then
$p\in\overline{\f^{-1}(I^{+}(\f(p)))}$ and according to
lemma \ref{CLOSURE} $I^{+}(p)\subseteq\f^{-1}(I^{+}(\f(p)))$
so that $\f(I^{+}(p))\subseteq I^{+}(\f(p))$.
As this holds for every $p\in V$ and $W$ is distinguishing,
proposition
\ref{SET2} ensures that $\f$ is a causal relation.\N
\begin{coro}
Let $W$ be future and past distinguishing. Then, a diffeomorphism
$\f : V\rightarrow W$ is a causal relation if and only if
$\f^{-1}(\B)$ is an achronal boundary for every achronal boundary
$\B\subset W$.
\end{coro}

All in all, the theorems, corollaries and 
propositions proved in this section \ref{sec:APP-CAUSA}, together with 
Propositions \ref{CONSERVATION} and \ref{CAUS} and Examples 6, 7, 9 and 11,
provide a sufficiently long list of causal objects and properties preserved
by causal relations.  Irrespective of the above comments on
the role which, for instance, future/past sets may play in causality theory, 
the mentioned list gives sufficient examples of nontrivial causal properties
shared by isocausal Lorentzian manifolds from what we conclude that isocausality
is actually isolating some essential information about the global causality of
the equivalence classes defined by $\sim$. On top of this, as we are going to
show in section \ref{subsect:CB}, isocausality is a useful tool in the study
of causal boundaries, and allows to generalize and improve the
causal diagrams of Penrose type.

\section{Causal extensions, causal diagrams and
causal boundary of spacetimes}\label{subsect:CB}
The idea of attaching a causal boundary to a spacetime $V$ was perhaps
first developed by Penrose \cite{CONF-BOUND,P1,P2} who used a {\em conformal}
embedding of $V$ into a larger Lorentzian manifold and defined the causal
boundary as the boundary of the embedded $V$ in the larger manifold. This idea
was subsequently refined by Geroch, Kronheimer and Penrose in \cite{GKP},
where a more general construction for such a boundary (which made use of no
embedding in principle) was performed with the only aid of the causal structure
of the spacetime under investigation---for {\em distinguishing} spacetimes.
Although the construction in \cite{GKP} yielded a satisfactory causal boundary
for many relevant spacetimes, it presents some difficulties with other,
causally worse-behaved, spacetimes.
One example is the Taub spacetime for which the causal boundary obtained
by this method does not match the knowledge obtainable using more
elementary means, see \cite{CAN-BIN}. Moreover, in order for the causal
boundary to be a Hausdorff topological space in this construction,
one has to provide an identification rule for the points in the boundary.
The original identification rule proposed in \cite{GKP} does not work
accurately in full generality, so that some alternative identification rules
and topological constructions were tried for spacetimes with good enough causal
properties, see \cite{RACZ,SZABADOS,HARRIS2}. Unfortunately, they eventually
turned out to be not as general as it was initially claimed
\cite{ZHIQUAN,HARRIS2}.
There are several other different ways of constructing a boundary (not
necessarily ``causal'') for Lorentzian manifolds, see
\cite{GEROCH,SCHMIDT,BUDIC,MEYER,SS}. Almost all
of them have failed to give a boundary with adequate topological properties for
some examples \cite{WALD,LIANG,HARRIS2}. This has led some researchers to the
opinion that not every distinguishing spacetime possesses a proper boundary.

Nevertheless, we would like to contribute to the subject with a new 
try which is a useful complement to the previous ones and may be 
helpful in several situations, although perhaps it does not solve all the 
difficulties just mentioned. As we have already shown, causal relationship 
generalizes 
---and in many cases is more useful and manageable than--- the
conformal relationship. Given that the Penrose conformal diagrams are
based on conformal relations,
we can try to generalize Penrose's ideas by using causal relations.
In this way we try, on one hand, to attach causal boundaries to general
spacetimes, and
on the other, to get some intuition and understanding of complicated
spacetimes
by analyzing the simpler ones to which they are isocausal.

To achieve these goals, we first of all need to include our spacetime
$V$ in a
larger one (such that the former has a boundary in the latter) but
{\em keeping the causal structure} of $V$. We do this as follows
(compare \cite{SS} and definition 3.1 in \cite{COND}).
\begin{defi}
An {\bf envelopment} of $V$ is an embedding $\Phi :V\rightarrow
\tilde{V}$ into
another connected manifold $\tilde{V}$ with $\Phi(V)\subset
\tilde{V}$.
A {\bf causal extension} of $V$ is any envelopment into another
Lorentzian
manifold $\tilde{V}$ such that $V\sim \Phi(V)$.
\label{caus-ext}
\end{defi}
Observe that, as is clear, a causal extension for $V$ is in fact a
causal extension for coset$(V)$, that is, for all $W$ such that $W\sim V$.
It must be remarked that the causal extension is different from the usual
extensions in which the (conformal) {\em metric} properties of $V$ are kept.
Here we only care about the causal structure of $V$, {\em in the 
sense of} definition \ref{STRUC}, which is at a more basic level.
Nevertheless, as is
clear any metric or conformal extension is in particular also a causal extension.
Of course, as is always
the case with extensions, the general causal extensions are not
unique, but this is
irrelevant for our purposes. Notice that any conformal embedding is in
fact
a causal extension of the type defined above {\em with the particular
choice}
that the causal equivalence between $V$ and $\Phi(V)$ is of conformal
type.
We drop here this condition and thereby we generalize the conformal
diagrams.
The more general diagrams constructed by means of causal extensions
will be called {\em causal diagrams}.
\subsubsection*{Example 12}
We saw in Example 3 that flat spacetime $\L$ and the static region of
de Sitter
spacetime $\fr{1}{4}\dS$ are causally equivalent:
$\L\sim\fr{1}{4}\dS$. Similarly,
we proved in Example 4 that $\RW_0\{\fr{3-n}{n-1}\}\sim\L$. It is completely
obvious that
the whole de Sitter spacetime $\dS$ is a causal extension of
$\fr{1}{4}\dS$,
hence $\dS$ is a causal extension also for $\L$ and $\RW_0\{\fr{3-n}{n-1}\}$.
Actually,
$\dS$ is a causal extension for coset$(\L)$. Note that flat spacetime
$\L$
is geodesically complete and therefore is not extendible in the usual
metric
way, but it is certainly extendible in the causal (including the
conformal) way.

With this causal extension for $\L$, all members of coset$(\L)$ have
a boundary
when seen as submanifolds of $\dS$. This boundary has a shape of type
``$>$'', and
it is formed by two null components, and one corner {\em which is a
$(n-2)$-sphere}
(the upper and lower corners
are {\em not} part of $\dS$ and therefore they are not part of the
boundary),
see figures \ref{Minkowski}, \ref{QUART} and \ref{STEADY}(c). They
correspond,
respectively,
to the horizon ($r\rightarrow \alpha$) of $\fr{1}{4}\dS$; to the
spacelike and
future null infinity and the past singularity in $\RW_0\{\fr{3-n}{n-1}\}$; and
to the spacelike and null infinity of $\L$. This is illuminating
in three respects:
\begin{itemize}
\item
firstly, because this boundary for coset$(\L)$ does not
distinguish between singularities, infinities or removable
singularities.
It only provides a shape and a {\em causal character} for the
boundary.
This is due to the fact that the specific metric properties have been
dismissed.
However, one can still recover the distinction between these types of
boundaries
by including endless curves as will be shown in subsection
\ref{asymptotia}.
\item
secondly, because the boundary found in a given causal extension
may not be what one expects to be the {\em entire} boundary of a given
spacetime. In this particular example, we can also perform another
causal extension which includes the upper and lower corners as part
of the boundary, i.e. future and past timelike infinity
for $\L$, as for instance the typical conformal embedding of $\L$ into the
Einstein universe $\E$ which is used traditionally to construct the
Penrose conformal diagram of $\L$ \cite{FF,CONF-BOUND,P2}. Observe
that then spacelike infinity becomes a point, while in the causal extension to
$\dS$ it is a $(n-2)$-sphere. This last possibility may be related to
the ideas developed by Friedrich in his treatment of conformal field
equations near the intersection of null and spacelike infinity, see e.g.
\cite{Fri} and references therein.
\item 
and thirdly, because the boundary built in some particular causal 
extensions may have different properties than those of other causal 
extensions, and they may even fail to have some reasonable or 
desirable features. For instance, it is well known that 
$\fr{1}{4}\dS$ is conformal to the region $T<-R$ of $\L$ 
(just take $T=-\alpha e^{-t/\alpha}/\sqrt{1-r^2/\alpha^2}$,
$R=re^{-t/\alpha}/\sqrt{1-r^2/\alpha^2}$ as the conformal mapping), 
so that a complete causal boundary for $\fr{1}{4}\dS$, and therefore
for $\L$ itself, can be seen as the union of the null hypersurface $T=-R$
with the corresponding part of past null infinity. The trouble here is that this
boundary is clearly distinct from the usual boundary obtained by the 
conformal embedding of $\L$ into $\E$. In the latter, $I^+(p\in {\cal J}^-)$
contains all of ${\cal J}^+$ except for part of one null generator, 
something which is untrue for the former.
\end{itemize}
Clearly, all this proves on one hand
that the causal boundaries found by these means are not unique nor 
with univocal properties, and on
the other that some of them might be more complete, and more appropriate,
causal boundaries than others. As a matter of fact, this is a 
circumstance 
also shared by the conformal boundary or GKP constructions. We refer the 
reader to the papers by Harris \cite{HARRIS1,HARRIS2} where the general
properties of ``reasonable'' causal boundary constructions for spacetimes
admitting a GKP causal boundary, as well as its 
possible universality, is considered. We will come back to this point 
later.

\subsubsection*{}
Despite all the problems mentioned in the previous discussion, we put 
forward the following definition of causal
boundary for Lorentzian manifolds based on the idea of isocausality.
\begin{defi}
Let $\tilde{V}$ be a causal extension of $V$ and $\d V$ the boundary
of $\Phi(V)$
in $\tilde{V}$. Then, $\d V$ is called the {\bf causal boundary} of
$V$
with respect to $\tilde{V}$. A causal boundary is said to be {\bf
complete} if
$\Phi(V)$ has compact closure in $\tilde V$.
\label{CAUSBOUND}
\end{defi}
Note that all the members in coset$(V)$ have the same causal boundary
{\em with respect to a given causal extension}. In principle, however,
the causal boundaries of coset$(V)$ depend on its causal extensions.
Moreover, the causal boundary may be empty.
\begin{prop}
If $V$ is compact, then its unique causal boundary is empty.
\end{prop}
\P A compact spacetime has no envelopment, because $\Phi(V)$ cannot
be a connected compact {\em open} proper subset of any $\tilde V$.\N

Of course, all compact spacetimes fail to be chronological \cite{BEE,FF,COND},
and so they are of little physical interest. It seems then reasonable to
assume that $V$ is distinguishing in order to attach a causal boundary
to $V$. Nevertheless, the causal boundary can still be
formed by discrete points (this is what one expects for the boundary of $\E$,
see figure \ref{EINSTEIN}), or be a set of any co-dimension in $\tilde V$.
It does not have to be connected either, as the de Sitter spacetime $\dS$ shows.
Despite all that, it appears to be natural that the causal properties of,
at least, the complete causal boundaries of coset$(V)$ for distinguishing $V$
will be in some sense the same, even though, as remarked before, 
there may be several different complete causal boundaries!

As we see, our proposal is just a refinement of the original Penrose's ideas
--mixed with some inspiration coming from the abstract boundary construction
of \cite{SS}-- by dropping the conformal property of the embedding:
we only require that the embedding be causal. Therefore, our definition covers
all the usual cases (such as Penrose's conformal embeddings and diagrams)
in which the causal boundary is built by means of a conformal embedding,
because a conformal relation is just a particular type of causal relation
(theorem \ref{CONF}). Similarly, the cases properly described by using 
the versions of \cite{GKP} which involve embeddings (there are 
non-embedding GKP constructions, see e.g. section 4 in \cite{HARRIS2})
should also be covered by
our definition due to the fundamental theorem \ref{KEY2} as the construction
in \cite{GKP} uses just (irreducible) future and past sets.
Our choice of {\em causal embedding} is motivated by the fact,
supported by the results found in section \ref{sec:APP-CAUSA},
that isocausality is a general concept keeping some important causal 
 properties of spacetimes. Thus, it seems sensible that
 if one wishes to maintain those causal properties of the
original Lorentzian manifold untouched, but without keeping the whole 
conformal structure, the general way of achieving this goal is by
using the isocausality concept.

We cannot claim at this stage that we have succeeded in attaching
a causal boundary to spacetimes where other techniques have failed, 
nor that we have improved the situation substantially. Nevertheless,
we can certainly attach a causal boundary,
proving also some of its relevant properties,
to some Lorentzian manifolds in a very simple manner. Perhaps the 
most noticeable property of our proposal is that the causal boundaries 
can be built by elementary means and quite easily for many, even 
complicated, spacetimes. Some explicit relevant cases are presented in the
next examples, including some cases where the Penrose conformal 
diagram cannot be drawn.

\subsubsection*{Example 13}
In this example we will construct a causal boundary for the
Schwarzschild
spacetime with negative mass. This spacetime is given in standard
spherical
coordinates by the line element ($-\infty<t<\infty , \, r>0,\ n\geq 4$)
\[
(\overline{\Sc},\T):d\tilde{s}^2=
\left(1+\fr{2M}{r^{n-3}}\right)dt^2-\left(1+\fr{2M}{r^{n-3}}\right)^{-1}dr^2-
r^2d\O^2_{n-2},
\]
where $M$ is a positive constant. In order to attach a causal
boundary to
$\bar{\Sc}$ we are going to put this spacetime in causal equivalence
with
another simpler spacetime for which a causal boundary is available.
Let us show that flat Minkowski spacetime with a timelike geodesic
removed does the job. To that end choose also spherical coordinates
for $\L$ and let
($-\infty<T<\infty , \, R>0$)
\[
(\L^*,\G):ds^2=dT^2-dR^2-R^2d{\bar\O}^2_{n-2}.
\]
 Notice that $\L^*$ is $\L$ with the line given by $R=0$ removed.
This is
in fact the line not covered by the coordinates just
introduced, and
defines the timelike geodesic previously mentioned. Thus, the base
manifold for
both $\bar{\Sc}$ and $\L^*$ is simply $\r\times(\r^{n-1}-\{O\})$
being ${O}$ a point of $\r^{n-1}$.

The needed diffeomorphisms are $\f:\L^*\rightarrow \overline{\Sc}$
defined simply
by $t=T$, $r=R$, and $\psi_{f}:\overline{\Sc}\rightarrow\L^*$ given by
$T=bt$, $R=f(r)$
where $b$ is a positive constant and the angular coordinates have been
identified as usual. The diagonal form of the tensors $\f^*\T$ and
$\psi^{*}_{f}\G$ in appropriate orthonormal bases read
\begin{eqnarray*}
\f^*\T=\mbox{diag}\left\{\left(1+\fr{2M}{R^{n-3}}\right),
-\left(1+\fr{2M}{R^{n-3}}\right)^{-1},\ -1,\dots,-1\right\},\\
\hspace{-1cm}
\psi^*_{f}\G=\mbox{diag}\left\{b^2\left(1+\fr{2M}{r^{n-3}}\right)^{-1},
-f^{'}(r)^2\left(1+\fr{2M}{r^{n-3}}\right),-\fr{f^2(r)}{r^2},\dots,
-\fr{f^2(r)}{r^2}\right\}.
\end{eqnarray*}
 From this formulae we readily see that $\f^*\T$ is a causal tensor
for every $R>0$ whereas $\psi^*_{f}\G\in\DP^+_2(\overline{\Sc})$ if the
following
restrictions are fulfilled
\[
b\geq f^{'}(r)\left(1+\fr{2M}{r^{n-3}}\right),\ b\geq
\left(1+\fr{2M}{r^{n-3}}\right)^{\fr{1}{2}}\fr{f(r)}{r}.
\]
These are achieved by for instance
$$
f(r)=\fr{r^{n-2}}{1+\left(r/M\right)^{n-3}}.
$$
Therefore we have proven that $\overline{\Sc}\sim\L^*$ and thus a
complete causal boundary for $\L^*$ will also be a complete causal boundary for
coset$(\L^*)\ni \overline{\Sc}$. A possible causal boundary for the former
consists on the usual causal boundary of flat spacetime $\L$ plus the removed
timelike geodesic. This last line corresponds to the curvature singularity of
$\overline{\Sc}$ located at $r=0$ which, in this particular causal 
extension, is timelike and represented by a point at each instant of time.

By a similar method one can prove that the inner part of
Reissner-Nordstr\"om spacetime, the one which contains the singularity and
is defined by $r<r_-$ (see Example 10), is also isocausal to $\L^*$.
Thus, the singularity there is also ``pointlike'' for these causal 
extensions.

At this stage, we may ask ourselves: how much the previous conclusions
depend on the particular causal extension used? It turns out that we 
can use here the powerful results in \cite{HARRIS1,HARRIS2} to say a word about 
how the causal boundary would look like for other causal extensions of
$\overline{\Sc}$. Since the GKP construction can be performed explicitly
for $\overline{\Sc}$ as well as for $\L^*$ (with a result 
similar to the one just obtained by our procedure), we can apply the
Theorem 3.6 in \cite{HARRIS2} to ensure that every other causal extension
will give a causal boundary with the same chronological properties as the one we
have constructed here. This also happens for some other
examples in this paper. Thus, the pointlike nature of the singularity 
in $\overline{\Sc}$, or in the inner part of Reissner-Nordstr\"om 
spacetime, seems firmly established. Perhaps the advantage of our method
lies on its generality and on the elementary means involved in our definition
which leads to a much easier and quicker construction than that
arising from the GKP definition, which is far more difficult to handle. 
In summary, it seems clear that the joint use of the GKP construction
with our definition in combination with the afore-mentioned results of
\cite{HARRIS2} can provide a very powerful machinery to deal with
the causal boundaries of very general spacetimes.

\subsubsection*{Example 14}
In this and the following examples we will construct a causal
boundary for some
anisotropic but spatially homogeneous ``Bianchi-I'' spacetimes
\cite{KRA},
including the
relevant cases of the Kasner Ricci-flat solutions, and the general
solution
for a comoving dust. All our considerations will be in arbitrary
dimension
$n$ in these examples, but we have kept the 4-dimensional terminology.
This kind of spacetimes has already been used 
in other causal boundary constructions \cite{GEROCH,HARRIS2}.

The general Bianchi-I Lorentzian manifold, denoted here by $\BI$, is
characterized
by having an Abelian group of motions of $n-1$ parameters acting
transitively
on spacelike hypersurfaces, the line-element taking the form
($j=1,\dots,n-1$)
\be
(\BI,\T) : d\tilde{s}^2=d\bar{t}^2-\sum_{j=1}^{n-1}
A^2_j(\bar{t})(d\bar{x}^j)^2,\
-\infty<\bar{x}^j<\infty
\label{BI}
\ee
where the $A_j(\bar{t})$ are arbitrary functions and the range of the
coordinate
$\bar{t}$ depends on their particular form.
We try to causally compare these spacetimes
with the general $n$-dimensional Robertson-Walker geometry with
flat slices $\RW_0\{a(t)\}$,
already studied in Example 4, and whose line-element in
Cartesian-like coordinates takes the form
\be
(\RW_0\{a(t)\},\G) : ds^2=dt^2-a^2(t)\sum_{j=1}^{n-1} (dx^j)^2,\
-\infty<x^j<\infty
\label{RW}
\ee
where $a(t)$ is the scale factor, which in particular defines the
range
of the time coordinate $t$. To start with, the diffeomorphism
$\f_f :\RW_0\rightarrow \BI$ will be chosen as
$(\bar{t},\bar{x}^j)=(f(t),x^j)$
where $f$ is a function to be determined. Then, the eigenvalues
of $\f_f^*\T$ with respect to $\G$ read
$\left\{f'^2(t),A_j(f(t))/a^2(t)\right\}$
so that criterion \ref{ORTHONORMAL} tells us
\be
\f^{*}\T\in\DP^+_2(\RW_0)\Longleftrightarrow a^2(t)f'^2(t)\geq
A^2_j(f(t)),\,\,
j=1,\dots,n-1.\label{der}
\ee
If these relations are fulfilled, then from corollary \ref{pull-back2}
$\RW_0\{a(t)\}\prec \BI$. The reciprocal
diffeomorphism $\psi_{\bar{f}} : \BI \rightarrow \RW_0$ will be taken
as
$(t,x^j)=(\bar{f}(\bar{t}),\bar{x}^j)$.
The eigenvalues of $\psi^*_{\bar{f}}\G$ with respect to $\T$ are
$$
\left\{ \bar{f}'^2(\bar t),\frac{a^2(\bar{f}(\bar{t}))}{A_j^2(\bar
t)}\right\}
$$
so that again criterion \ref{ORTHONORMAL} provides
\be
\psi_{\bar{f}}^*\G\in\DP^+_2(\BI)\Longleftrightarrow
\frac{\bar{f}'^2(\bar t)}{a^2(\bar{f}(\bar t))}\geq
\frac{1}{A^2_j(\bar t)}, \,\,
j=1,\dots,n-1.\label{izq}
\ee
If (\ref{izq}) are complied, then $\BI\prec \RW_0\{a(t)\}$
from corollary \ref{pull-back2}.
When both (\ref{der}) and (\ref{izq}) are satisfied, then $\BI\sim
\RW_0\{a(t)\}$.
Of course, these are not the only possibilities that make the causal
relationship
between $\BI$ and $\RW_0$ possible, as we have just tried some very
particular
diffeomorphisms, but they will be enough to prove the causal
equivalence
of some important subcases of the Bianchi-I spacetimes with the
simpler
and easier to handle Robertson-Walker Lorentzian manifolds.

Let us start by restricting the $\BI$ to be a generalized Kasner
spacetime,
denoted by $\K\{p_j\}$ and defined by $A_j(\bar t)=\bar{t}^{p_j}$ for
some constants $p_j$ (called the Kasner exponents), and $\bar
t\in(0,\infty)$.
The condition (\ref{der}) is fulfilled if we choose $f(t)=e^t$
(with $-\infty<t<\infty$) and
$a(t)=B+e^{-kt},\ k=\mbox{max}\{1-p_j\},\ B>1$
whenever all the $p_j$ are such that $p_j\leq 1$. To establish the
causal
equivalence in this case, it only remains to find an $\bar f$ such
that (\ref{izq})
holds too. For instance, the function
$h(\bar{t})=\fr{Q}{1+\log^2 \bar{t}}+\bar{t}^q$ satisfies
$h(\bar{t})\geq \bar{t}^{1-p_j}$ for all $\bar{t}\in(0,\infty)$
if $q=\mbox{max}\{1-p_j\}>0$ and $Q\geq 1$.
Thus, the solution of the differential equation
\be
\fr{\bar{f}^{'}(\bar{t})}{a(\bar{f}(\bar{t}))}=
\fr{1}{\bar{t}}\left(\fr{Q}{1+\log^2 \bar{t}}+\bar{t}^q\right)
\label{EQDIFF1}
\ee
will provide the required $\bar f$. This equation can be solved and
the general
solution reads
$$
\bar{f}(\bar{t})=
\fr{1}{k}\log\left\{\fr{1}{B}\left[A\exp\left[ kB\left(Q\arctan (\log
\bar{t})
+\fr{\bar{t}^q}{q}\right)\right]-1\right]\right\},
$$
where $A$ is an integration constant which must be arranged in such a
way that
the image of $\bar{f}(\bar{t})$ for $\bar{t}\in(0,\infty)$ covers the
whole real
line. This is accomplished by taking $A=e^{\fr{kBQ\pi}{2}}$.

Thus, we have proved that $\K\{p_j< 1\}\sim\RW_0\{B+e^{-kt}\}$ if
$k=\mbox{max}\{1-p_j\},\ B>1$. This case includes the proper Kasner
spacetimes,
which are the particular cases with
\be
\sum_{j=1}^{n-1}p_j=\sum_{j=1}^{n-1}p_j^2=1, \label{exponents}
\ee
and then the
manifolds are Ricci flat (i.e., solutions of the vacuum Einstein
field equations.)
 From the constraints (\ref{exponents}) we clearly have $|p_j|\leq 1$
so that
if any of the $p_j$ is one then the rest of the $p_j$ must vanish.
This particular simple spacetime with $p_1=1$ and $p_k=0$ ($k=2,\dots,n-1$)
will be dealt with presently as a particular subcase of the general case with
$p_1=1$, see Example 16.

\begin{figure}[h]
\epsfxsize=10cm
\hspace{4cm}\epsfbox{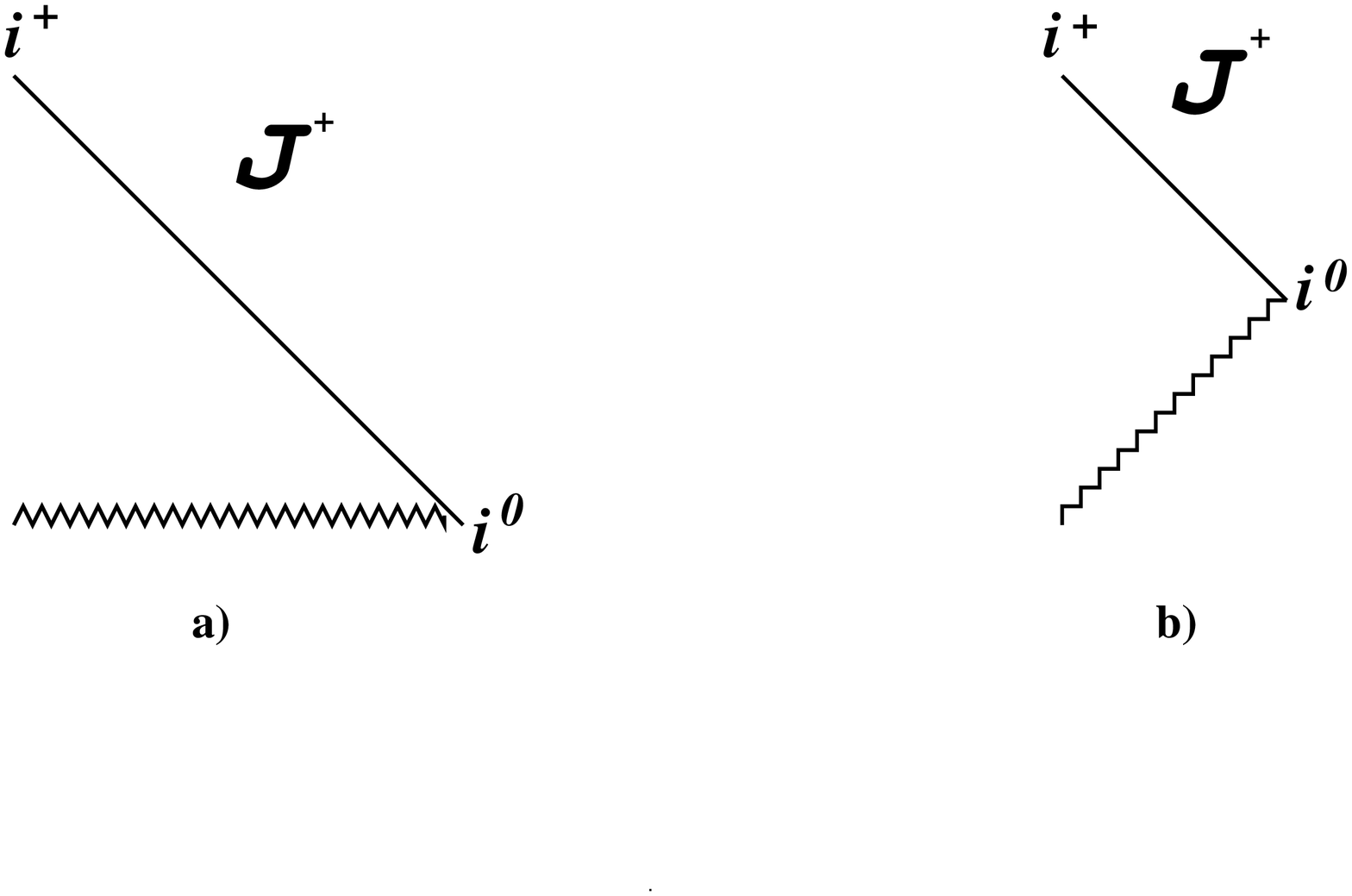}
\vspace{-0.3cm}
\caption{\label{STAT} These are the causal diagrams for the Kasner-type
spacetimes. A 3-dimensional version could also be easily drawn by
letting these figures become surfaces of revolution around their vertical
left ends. We specifically show the causal boundary for Kasner vacuum spacetimes
in figure (a) and for the anisotropic Bianchi I dust models (figure (a) if 
$p_{j}<1$ and figure (b) if $p_{j}>1$).
We see that in each case the causal boundary has two components of
different causal character, and that the singularity is of null type in the
second case.}
\end{figure}

Once the causal equivalence has been established, we can work with
the easier
and simpler $\RW_0\{B+e^{-kt}\}$ and try to attach a causal boundary
to the
class coset$(\K\{p_j< 1\})$ by attaching it to
$\RW_0\{B+e^{-kt}\}\in$ coset$(\K\{p_j< 1\})$. This is a rather
simple task
since, for instance, $\RW_0\{B+e^{-kt}\}$ can be written in explicitly
conformally flat form by means of the coordinate transformation
$\t=\fr{1}{kB}\log(1+Be^{kt})$
\be
ds^2=\fr{B^2e^{2kB\t}}{(e^{kB\t}-1)^2}(d\t^2-\sum_{j=1}^{n-1}
(dx^j)^2),
\label{EQ-TO-K}
\ee
where $\t\in(0,\infty)$. In consequence, the whole flat spacetime
$\L$ (for
$\t\in(-\infty,\infty)$) is clearly a causal extension of
$\RW_0\{B+e^{-kt}\}$
according to definition \ref{caus-ext}, and then $\dS$ or $\E$ are yet
larger causal extensions. From this we deduce that
a causal boundary for coset$(\K\{p_j< 1\})$ is the
hypersurface constituted by one spacelike component (the hypersurface
$\S$ given by $\t=0$) representing the past singularity, one null
component
which represents the null infinity ${\cal J}^+$ of flat spacetime,
their intersection at $i^0$ and the point ${\it i}^+$,
which correspond respectively to
the spacelike and future timelike infinity of Minkowski spacetime.
This was to be expected and can be represented in the schematical
causal diagram of figure \ref{STAT}-(a). Notice that this contradicts a 
result found in \cite{HARRIS2} (p.598), but it seems that correct application 
of Proposition 5.2 in that paper would lead to the same conclusion as 
ours.

If on the other hand we assume that every Kasner exponent $p_j$ is
greater than
one, then in the very same way we can prove that $\K\{p_j>1\}$ is
isocausal with
the spacetime with line-element (\ref{EQ-TO-K}) where now
$-\infty<\tau<0$.
Thus we get for this spacetime the diagram of figure \ref{STAT}-(a)
but turned
upside-down, with the roles of future and past interchanged. The null
component of the causal boundary corresponds now to the singularity 
and the spacelike component to future infinity, in a behaviour
analogous to that of $\RW\{\g\in(-1,\fr{3-n}{n-1})\}$ shown in figure
\ref{STEADY} (b). Compare with \cite{HARRIS2}.

\subsubsection*{Example 15}
It is interesting to repeat the previous calculations for the $\BI$
model
with $A_j(\bar{t})=\bar{t}^{p_j}(\bar{t}+t_0)^{\fr{2}{3}-p_j}$,
$\bar{t}\in(0,\infty)$, where $t_0$ a positive constant. This
spacetime has
physical interest as it provides (in $n=4$) the general solution of
Einstein's equations for Bianchi-I pressure-free perfect fluids
whenever the
conditions (\ref{exponents}) are assumed. We will nonetheless allow
for
other values of the $p_j$. For instance, assume that all the exponents
are such that $p_j>1$. The conditions (\ref{der}) hold then if, for
example,
$f(t)=e^t$ and $a(t)=A/\cosh\mu t$ for several suitable choices of
the positive constants $A$ and $\mu$. Similarly, conditions
(\ref{izq}) are
then satisfied by choosing $f(\bar{t})=B\log\bar{t}$ for some
suitable $B$
and rearranging, if needed, $A$ and $\mu$ in order to comply with all
the
inequalities. Thus, these $\BI$ models are isocausal to
$\RW_0\{\fr{A}{\cosh\mu t}\}$ which may be rewritten in an explicitly
conformally flat form by means of the coordinate change
$\cosh(\mu t)\, dt=A d\t$, leading to
$$
d\tilde{s}^2=\fr{A^2}{1+\mu^2A^2\t^2}
\left(d\t^2-\sum_{j=1}^{n-1}(dx^j)^2\right),\
-\infty<\t<\infty .
$$
We conclude that these $\BI$ spacetimes are isocausal to $\L$ and thus
we can attach a causal boundary to them similar to that of $\L$, with
two
null components and three corners, see figure \ref{STAT} (b).
The singularity turns out to be null again, and the future infinity
is now
of Minkowskian type, with a null component and a point. This is
similar
to the behaviour of $\RW_0\{\fr{3-n}{n-1}\}$ shown in figure \ref{STEADY} (c).

If all the exponents are lower than one, it is easy to check that the
$f(t)=e^t$
and the scale factor $a(t)=Q_1+e^{-\alpha t}$ for some adequate $Q_1$
and $\alpha$
provide an isocausal $\RW_0$ spacetime. The causal boundary in this
case, which is the one of physical relevance in $n=4$,
turns out to be equivalent to that of Ricci-flat Kasner spacetime, figure
\ref{STAT} (a).
\subsubsection*{Example 16}
 In our study of the causal boundary of $\K\{p_j\}$ we have
restricted ourselves
to the situation in which all the exponents are either greater, or
lower, than one.
It is not difficult to see that the techniques used in the Example 14
do not work
for the mixed case with some $p_i\geq 1$ and some other $p_j\leq 1$
since the
diffeomorphisms constructed there fail to be causal relations for the
whole
Lorentzian manifolds. Now we are going to prove that
$\K\{p_1=1,p_k<1\}$
($k=2,\dots ,n-1$)
is isocausal with a precise subregion of flat Minkowski spacetime,
and thereby we are going to construct the causal boundary for these
spacetimes.
The computations work also for the case of $\K\{p_1=1,p_k>1\}$ with
slight changes
and a different region of $\L$.

\begin{figure}[h]
\epsfxsize=6cm
\hspace{5cm}\epsfbox{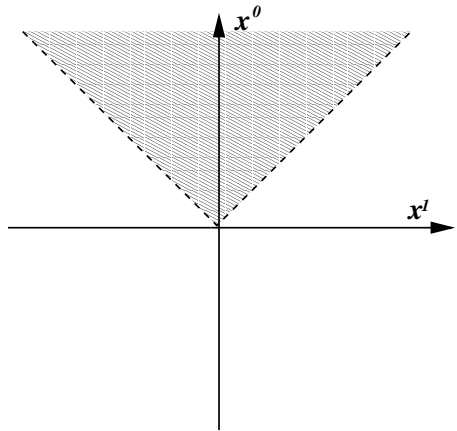}
\caption{The shaded region of this picture represents the spacetime $\L_+$.  We
appreciate clearly that this is a submanifold of the full Minkowski
spacetime $\L$ which metrically extends $\L_+$.  In a rather similar
way the picture for $\L_-$ is obtained from this figure by just
turning it upside-down (region $x^0<-|x^1|$).}
\label{KASNERMIX}
\end{figure}

To illustrate why one should expect a more complex causal boundary
for these
cases, and to get an idea of which type of causal boundary, consider
first
the extreme case $p_1=1$ such that (\ref{exponents})
hold. Then, the rest of the exponents vanish and therefore we have
$$
p_1=1,\ p_k=0,\ \,\, \forall\  k=2,\dots,n-1 .
$$
As is well-known, this special Kasner spacetime $\K\{p_1=1,p_k=0\}$
is just a region of flat spacetime $\L$. To see it, simply perform
the coordinate
change $\{\bar{t},\bar{x}^j\}\rightarrow \{x^{\alpha}\}$ defined by
$x^0=\bar{t}\cosh \bar{x}^1$, $x^1=\bar{t}\sinh\bar{x}^1$, and
$x^k=\bar{x}^k$ for $k=2,\dots,n-1$. This
change is well-defined for $x^0>|x^1|$ and the line-element takes the
manifestly
flat form
\begin{equation}
\left(\L_+,\G\right)\, : \, \, \, \,
ds^2=(dx^0)^2-\sum_{j=1}^{n-1}(dx^j)^2, \hspace{1cm} x^0> |x^1| \, .
\label{L+}
\end{equation}
 From now on, we are going to call $\L_+$ the region of flat
spacetime defined by
$x^0>|x^1|$ (see figure \ref{KASNERMIX}). Hence, we have proved that
$\K\{p_1=1,p_k=0\}=\L_+$. In particular this means that the
whole flat spacetime $\L$, and its causal extensions, are causal
extensions for
$\K\{p_1=1,p_k=0\}$ so that a complete causal boundary for
$\K\{p_1=1,p_k=0\}$
has now two null components to the past given by $0<x^0=\pm x^1$, a
corner at their
intersection, plus other null component to the future, corners
at the intersection of this with the previous ones, and the corner at
timelike infinity of $\L$. This is a little bit more complicated
structure,
and we claim that this is the type of boundary that other
$\K\{p_1=1,p_k<1\}$
will have.

Bearing this goal in mind, we are going to prove that, in fact,
$\K\{p_1=1,p_k<1\}\in$ coset$(\L_+)$. Consider first the
diffeomorphisms
$\psi_f : \K\{p_1=1,p_k<1\}\rightarrow \L_+$ defined by
$$
x^0=f(\bar t)\cosh (a \bar{x}), \,\,\, x^1=f(\bar t)\sinh (a
\bar{x}), \,\,\,
x^k=\bar{x}^k .
$$
Then, the eigenvalues of $\psi^*_f\G$ with respect to $\T$ are
$\{f'^2,a^2f^2/\bar{t}^2,1/\bar{t}^{2p_k}\}$ so that we have
$$
\psi^{*}_f\G\in\DP_2^+(\K\{p_1=1,p_k\})\Longleftrightarrow
f^{'}(\bar t)\geq\fr{a\, |f(\bar t)|}{\bar t},\
f^{'}(\bar t)\geq\fr{1}{\bar{t}^{p_k}}.
$$
Hence, for instance every non-decreasing solution of the differential
equation
$$
f^{'}=\fr{a}{\bar t}f +\sum_{k=2}^{n-1}\bar{t}^{-p_k}
$$
will comply with the previous conditions. The general solution of this
differential equation is given by ($C$ is an arbitrary integration
constant)
\[
f(\bar{t})=C\bar{t}^{a}+\sum_{k=2}^{n-1}\fr{\bar{t}^{1-p_k}}{1-p_k-a}
\]
which is easily seen to define a true diffeomorphism $\psi_f$
as long as $0<a<$ min$(1-p_k)$.

\begin{figure}[h]
\epsfxsize=12cm
\hspace{3cm}\epsfbox{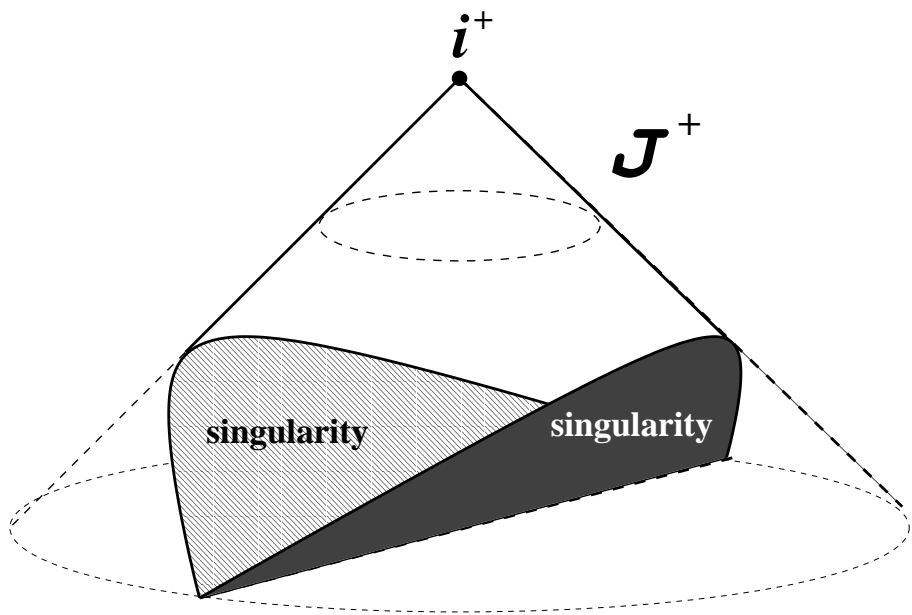}
\caption{This is the causal diagram for all Kasner-like spacetimes
with $p_1=1$ and $p_k<1$. The spacetime is the part above the singularity and
below the future infinity ${\cal J}^+$. As we see, a 3-dimensional
figure is needed here to account for the basic properties of the
causal boundary.
The past singularity is of null type with two branches. The spacetime has
no past particle horizon in the direction defined by the null
generator of this singularity. However, there
are past particle horizons in every other direction. The future infinity
has the usual structure of Minkowski spacetime.
In the case $p_k>1$, the causal diagram is the
one shown here but turned upside down, with the roles of the
singularity
and infinity interchanged, so that the singularity is also in the
past and now there are future particle horizons, but no past ones.
}
\label{Trid}
\end{figure}

 For the converse causal relation, let $t$ denote
$\sqrt{(x^0)^2-(x^1)^2}$
and choose a diffeomorphism $\f_h:\L_+\rightarrow \K\{p_1=1,p_k\}$ of
type
$$
\bar{t}=h(t), \,\, \bar{x}=\fr{b}{2}\log
\left(\fr{x^0+x^1}{x^0-x^1}\right),
\, \, \, \bar{x}^k=x^k .
$$
In this case, if one computes $\f_h^*\T$ there appear crossed terms
in the
given coordinates, however
by considering the following orthonormal basis in $\L_+$
$$
\left\{\fr{1}{t}(x^0dx^0-x^1dx^1),\fr{1}{t}(x^1dx^0-x^0dx^1),
dx^k\right\}
$$
it is easily seen that the eigenvalues of $\f^*_h\T$ with respect to
$\G$ are
given by $\{h'^2,b^2h^2/t^2,h^{2p_k}\}$ so that
$$
\f^*_h\T\in\DP_2^+(\L_+) \, \Longleftrightarrow \,
h'\geq\fr{|h|b}{t},\ h'\geq h^{p_k}.
$$
A possible function $h$ satisfying the previous requirements is given
by
$h(t)=At^c+Bt^d$ for appropriate values of $A$ and $B$ as long as the
parameters
$c$ and $d$ obey
\[
c\geq\mbox{max}\left\{\fr{1}{1-p_k}\right\}\geq
\mbox{min}\left\{\fr{1}{1-p_k}\right\}\geq d\geq b > 1 .
\]
Under these assumptions, $h(t)$ is a diffeomorphism of $\r^{+}$ into
itself
from what we
finally obtain the desired result $\L_+\sim\K\{p_1=1,p_k<1\}$.
Thus, with the causal equivalence just constructed, the causal
diagram for $\K\{p_1=1,p_k<1\}$ can be easily constructed,
and we can also attach a complete causal boundary to them, which is the
causal boundary of coset$(\L_+)$ previously mentioned.
This is shown in figure \ref{Trid}.

Similarly, one can prove along the same lines that
$\K\{p_1=1,p_k>1\}\in$ coset$(\L_-)$ where now $\L_-$ is the region
of flat
spacetime defined by $x^0<-|x^1|$ (see figure \ref{KASNERMIX}).
In this case the causal diagram is the time reversal of
that in figure \ref{Trid}, but now of course with the roles of the
singularity and
infinity interchanged so that the singularity is still in the past.

\subsection{Identifying the parts of a causal
boundary}\label{asymptotia}
We can use the ideas developed so far to try to identify the various
parts of a causal boundary. As we have seen, the causal boundaries
can be endowed with causal properties
(we can say if they have null components, or
spacelike ones, etc), but we are not able to say if these components
are part
of the singularity, or of infinity. To do that, of course, we must
use at least
part of the metric properties of $V$: essentially those
characterizing the
completeness or not of causal curves (see e.g. \cite{FF,Wald,COND} for
the
definition of completeness). Namely
\begin{defi}
Let $\d V$ be the causal boundary of $V$ with respect to the causal
extension
$\tilde V$. A point $\tilde p\in\d V$ is said to belong to:
\begin{enumerate}
\item a {\bf singularity set} ${\cal S}\subseteq \d V$ if it is the
endpoint in
$\tilde V$ of a curve which is endless and incomplete within $V$.
\item {\bf future infinity} ${\cal J^+}\subseteq \d V$ if it is the
endpoint
in $\tilde V$ of a causal curve which is complete to the future in
$V$.
And similarly for the past infinity ${\cal J^-}$.
\item {\bf spacelike infinity} $i^0\subseteq \d V$ if it is the
endpoint
in $\tilde V$ of a spacelike curve which is complete in $V$.
\end{enumerate}
\end{defi}
The unions of all past and future infinities will be also termed as
{\em causal infinity}. In principle, there is no reason to believe
that all points in a causal boundary belong to one of the possibilities of
the previous definition, nor that the different possibilities are
disjoint in general.

\subsubsection*{Example 17}
Coming back to the Example 12, where we saw that $\dS$ was a causal
extension
for coset$(\L)\supset\left\{\L,\fr{1}{4}\dS,\RW_0\{\fr{3-n}{n-1}\}\right\}$,
we can now
easily identify the different parts of the causal boundary. For $\L$,
which
is b-complete (all curves are complete), the upper (lower) null
component of the
boundary is future (past) null infinity, and the corner is spacelike
infinity.
For $\fr{1}{4}\dS$, the whole causal boundary is a singularity. Of
course, this is
a {\em removable} singularity \cite{COND}, as $\fr{1}{4}\dS$ can be
metrically
extended to the whole of $\dS$, but this is another matter. Finally,
for
$\RW_0\{\fr{3-n}{n-1}\}$ the future null component of the causal boundary
represents
future infinity, the corner is spacelike infinity, and the  past null
component
is the ``big-bang'' singularity. Note that in this case the
singularity is
essential (irremovable by metric extensions \cite{COND}). A similar
behaviour
is the one found for the Bianchi-I dust spacetimes in figure
\ref{STAT} (b).

\subsubsection*{}
The above considerations allow us to put forward a tentative
characterization of {\it causally asymptotically equivalent
spacetimes} at a
level which does not use the whole information contained in the
Lorentzian
metric. This information might be included in a subsequent step if
one wishes
to define asymptotically flat, or asymptotically $\dS,\ \AdS$, etc,
spacetimes.
We are now just caring about the causal properties of the asymptotic
structure, only distinguishing between infinities and singularities.
The idea
is to say that two spacetimes have the same asymptotic properties
from the
causal point of view if there are arbitrarily small neighbourhoods of
the
relevant parts of their causal boundaries which are isocausal.
To that end, we need to know what is a neighbourhood of a causal
boundary and
its parts.
\begin{defi}
An open set $\zeta\subset V$ is called a {\bf neighbourhood of}
\begin{enumerate}
\item the causal boundary
of $V$ if $\zeta\cap \g\neq \emptyset$ for all
endless causal curves $\g$;
\item a singularity set ${\cal S}$ if $\zeta\cap \g\neq \emptyset$
for all endless curves $\g$ which are incomplete towards ${\cal S}$;
\item causal (future, past) infinity if $\zeta\cap \g\neq \emptyset$
for all complete (complete to the future, to the past) causal curves
$\g$.
\end{enumerate}
\label{as1}
\end{defi}
Let us remark that we do not need to use any causal extension in the
preceding
definition. Only the properties of $V$ are required.
Then we introduce the following definition, which might require some
refinement.
\begin{defi}
$W$ is said to be causally asymptotically like $V$ if any two
neighbourhoods
of their causal infinities $\zeta\subset V$ and $\tilde\zeta\subset
W$ contain
corresponding neighbourhoods $\zeta'\subset\zeta$ and
$\tilde\zeta'\subset\tilde\zeta$ of the causal infinities such that
$\zeta'\sim\tilde\zeta'$.
\label{as2}
\end{defi}
Similar definitions can be given for $W$ having causally the
singularity structure of $V$, or the causal boundary of $V$,
replacing in the given definition the neighbourhoods of the causal
infinity by those of the singularity and of the causal boundary,
respectively.

In the previous Example 17, it is easy to see that $\RW_0\{\fr{3-n}{n-1}\}$ is
future
asymptotically ``flat'' (that is, future asymptotically equivalent to
$\L$)
from the causal point of view, while $\fr{1}{4}\dS$ has a
past-singularity
of the type of $\RW_0\{\fr{3-n}{n-1}\}$ (of course, removable!). A more
interesting
case arises from Example 9, where we proved that $\Sc_a\sim\L_a$ for
all $a>c_M$.
This clearly means, according to definitions \ref{as1} and \ref{as2},
that the causal infinities for flat and outer Schwarzschild
spacetimes are causally equivalent.

\section{Conclusions}

In this work a new tool for the causal analysis of Lorentzian
manifolds has been defined and developed. The most remarkable of its
properties are: (i) that the causality constraints are kept in 
isocausal spacetimes; (ii) the precise relationships 
between some causal objects---like achronal boundaries,
and future/past sets, curves, or tensors--; (iii) the generalization it
provides of conformal relations; and (iv) the refinement of the
classical causality conditions, which can also be considered in an
abstract way. Furthermore, the whole idea allows us to undertake the
study of global causal properties of spacetimes with a high degree of
generality, for one can get a first impression of the properties of
general spacetimes by studying other spacetimes which are simpler but
nevertheless isocausal to the former. In particular we can draw causal diagrams,
which are clear generalizations of the Penrose conformal diagrams, 
and obtain causal boundaries for general spacetimes by using quite
simple and elementary means.

Several questions remain open, as for example the need for
improvements and some general results on the criteria used to discard
or to prove the possible causal relation between given spacetimes,
the existence of upper and lower bounds for causally ordered sequences
of equivalence classes of isocausal spacetimes, the precise extent to 
which isocausal spacetimes can be thought of as sharing the same 
causal properties, and the intrinsic or uniqueness properties of the 
different causal boundaries constructed by means of causal extensions,
among others.

\section*{Acknowledgements}
We are grateful to Marc Mars and Ra\"ul Vera for reading the
manuscript and for many useful comments. We acknowledge also the anonymous
referees for their constructive criticism, the corrections and 
amendments suggested, and the improvements they impelled.
This research has been carried out under the research project UPV
172.310-G02/99 of the University of the Basque Country.

\end{document}